%% file: main.tex
\documentclass[11pt,a4paper,twoside]{article}
\usepackage[T1]{fontenc}
\usepackage[utf8]{inputenc}
\usepackage{amssymb}
\input{preamble.tex}

\begin{document}
% \title{The Boltzmann Equation for Nearly Incompressible Flows: A High-Order Arbitrary Lagrangian-Eulerian Discontinuous Galerkin Method}
% \title{Simulating Nearly Incompressible Flows with a High-Order Arbitrary Lagrangian-Eulerian Discontinuous Galerkin Method for the Boltzmann Equation}
\title{A High-Order Arbitrary Lagrangian-Eulerian Discontinuous Galerkin Method for the Boltzmann Equation in Nearly Incompressible Flows}
% \title{High-Order Arbitrary Lagrangian-Eulerian Discontinuous Galerkin Method for the Boltzmann Equations for Nearly Incompressible Flows}
\author[1]{A.~Aygun\footnote{Corresponding author. E-mail address: atakana@metu.edu.tr} }
\author[1]{O.~Ata}
\author[2]{T.~Warburton}
\author[1]{A.~Karakus}

\affil[1]{\textit{\small{Department of Mechanical Engineering, Middle East Technical University, Ankara, Turkey 06800}}}
\affil[2]{\textit{\small{Department of Mathematics, Virginia Tech, Blacksburg, VA, 24061 USA}}}

\renewcommand\Authands{ and }
\date{\vspace{-5ex}}
\maketitle

\begin{abstract}
We propose the arbitrary Lagrangian-Eulerian (ALE) form of the Galerkin-Boltzmann formulation for the simulation of nearly incompressible flows with moving boundaries. The continuous Boltzmann equations are mapped to a reference state to compensate the mesh motion with an advection term. The resulting system is discretized in space using the discontinuous Galerkin method on unstructured meshes. A semi-analytic Runge-Kutta time discretization is used to overcome the stiffness introduced by the continuous Boltzmann equations. The well-known geometric conservation law is shown to be satisfied by the time and space discretizations and consistent update of geometric factors of the discretization. The implementation is on the GPU accelerated kernel library libParanumal and validated by a free stream preservation and moving Taylor-Green vortex test cases. Then, the capabilities are shown using a plunging symmetric airfoil in two-dimensions and moving carangiform fish in three-dimensions using perfectly matched layers.

\end{abstract}

% \input{nomen.tex}
% \printnomenclature

\section{Introduction}
\input{introduction}

\section{Galerkin-Boltzmann Formulation in the ALE Form}
\input{methodology}

\section{Spatial Discretization}
\input{dg_discretization}

\section{Temporal Discretization}
\input{time_disc}

\section{Mesh Motion and Geometric Conservation Law}
\input{mesh_motion}

\section{GPU Implementation}
\input{GPU_implementation}

\section{Results}
\input{results}

\section{Conclusion and Future Work}
\input{conclusion}

\section*{Acknowledgments}
The numerical calculations reported in this paper were fully performed using the EuroHPC Joint Undertaking (EuroHPC JU) supercomputer MareNostrum 5, hosted by the Barcelona Supercomputing Center (BSC). Access to MareNostrum 5 was provided through a national access call coordinated by the Scientific and Technological Research Council of Turkey (TÜBİTAK).

\bibliographystyle{ieeetr}
\bibliography{main}

\end{document}

%% file: preamble.tex
\usepackage{lipsum}
\usepackage{authblk}
\usepackage[top=3cm, bottom=3cm, left=2cm, right=2cm]{geometry}
\usepackage{fancyhdr}
\usepackage{adjustbox}
\usepackage[sort&compress, numbers, compress ]{natbib}
\usepackage[colorlinks,bookmarksopen,bookmarksnumbered,citecolor=red,urlcolor=red]{hyperref}
\usepackage{algorithm}
\usepackage{algorithmic}
\usepackage{boxedminipage}
\usepackage{comment}
\usepackage{eqparbox}
\usepackage{multirow}
\usepackage{multicol}
\usepackage{caption}
\usepackage{subcaption}

\usepackage{nomencl}
\makenomenclature

\usepackage{amsmath}
\usepackage{amsfonts}

\usepackage{graphicx}

\usepackage{tikz}

\newlength{\commentindent}
\usepackage{array}
\newcolumntype{R}[1]{>{\raggedleft\let\newline\\\arraybackslash\hspace{0pt}}m{#1}}

\newcommand\BibTeX{{\rmfamily B\kern-.05em \textsc{i\kern-.025em b}\kern-.08em
T\kern-.1667em\lower.7ex\hbox{E}\kern-.125emX}}

\let\OLDthebibliography\thebibliography
\renewcommand\thebibliography[1]{
	\OLDthebibliography{#1}
	\setlength{\parskip}{0pt}
	\setlength{\itemsep}{0pt plus 0.3ex}
}

\renewenvironment{abstract}{%
	\begin{center}
		\begin{adjustbox}{minipage=0.90\textwidth,center} 
			\rule{\textwidth}{0.5pt}}
		{\par\noindent\rule{\textwidth}{0.5pt}  
		\end{adjustbox}
	\end{center} 
}

\makeatletter

\renewcommand\@maketitle{%
	\begin{adjustbox}{minipage=0.90\textwidth,center}
		\begin{center}
			\vskip 1em
			\let\footnote\thanks 
			{\LARGE \@title \par }
			\vskip 1.0em
			{\large \@author \par}
		\end{center}
	\end{adjustbox}
	\vskip 0.0em \par
}
\makeatother
\usepackage{etoolbox}
\renewcommand\nomgroup[1]{%
  \item[\bfseries
  \ifstrequal{#1}{G}{Greek Symbols}{%
  \ifstrequal{#1}{S}{Symbols}{%
  \ifstrequal{#1}{A}{Abbreviations}{%
  \ifstrequal{#1}{B}{Subscripts}{%
  \ifstrequal{#1}{P}{Superscripts}{
  }}}}}]}

%% file: introduction.tex
\label{sec:Introduction}
Simulating physical systems governed by partial differential equations (PDEs) on moving domains is a fundamental challenge in computational science and engineering. Applications such as fluid-structure interaction (FSI), arterial blood flow \citep{bazilevs2006bloodFsi}, rotating turbines \citep{horvath_conforming_2022}, moving airfoils \citep{persson_discontinuous_2009}, free surface \citep{fu_arbitrary_2020}, and multi-material flows \citep{zhang_bound-preserving_2024} require robust and accurate numerical methods capable of handling domain deformation over time. The numerical methods for moving domains can be classified into boundary fitted (interface tracking) and non-boundary fitted (interface capturing) methods. Non-boundary-fitted methods, including fictitious domain method \citep{glowinski2001fictitious},  the extended finite element method (XFEM) \cite{moes1999finite}, and immersed boundary methods \cite{peskin2002immersed} embed the moving geometry into a fixed or adaptively refined background mesh. These methods avoid mesh deformation by allowing the geometry to intersect arbitrary elements, simplifying mesh generation, but they have some difficulty in modifying for high-order accuracy. In boundary fitted methods, the computational mesh conforms to the evolving geometry of the physical domain. The arbitrary Lagrangian–Eulerian (ALE) method is one of the most popular methods in this class, where the mesh moves with a prescribed velocity decoupled from the material flow. An Eulerian framework uses a fixed mesh in space and struggles with moving boundaries and interfaces. A Lagrangian framework moves the mesh with the material but suffers from mesh tangling and distorted elements. The ALE method was developed to offer a more general formulation to overcome the difficulties of the Eulerian and Lagrangian descriptions. It was first developed in \cite{hirt1974arbitrary} with a finite difference discretization by choosing the mesh velocity independently from the material velocity, and later developed for the finite element method \citep{hughes_lagrangian-eulerian_1981}.

The integration of high-order spatial discretizations, such as the discontinuous Galerkin (DG) and spectral element methods (SEM), into the ALE framework enables the resolution of complex physical phenomena with minimal numerical diffusion. These benefits are especially valuable in simulations involving moving domains, where capturing complex features near deforming boundaries or interfaces is critical. A numerous successful contributions had been made throughout the years with high-order methods \cite{lomtev_discontinuous_1999, etienne_perspective_2009, persson_discontinuous_2009, minoli_discontinuous_2011, klingenberg_arbitrary_2017, fu_arbitrary_2019}. In these formulations, the physical moving domain is mapped to a fixed reference frame, allowing the conservation laws to be solved in a consistent coordinate system where the mesh velocity and deformation Jacobian are evaluated with the same high-order precision as the solution variables. To ensure that these high-order ALE schemes maintain their accuracy and stability during mesh motion, specific attention is given to the discrete geometric conservation law (GCL). The GCL provides a framework for ensuring that the numerical fluxes and volume changes are integrated in a manner that preserves constant solutions, such as uniform flow, across a deforming mesh \cite{thomas_geometric_1979}. Various successful strategies have been established to enforce this consistency in high-order methods focusing on the time evolution of the Jacobian \cite{persson_discontinuous_2009}, evaluating different ways of the mesh velocity \cite{etienne_perspective_2009} or constructing the discrete operators to satisfy GCL automatically \cite{minoli_discontinuous_2011, fehn_high-order_2021}.

Beyond their theoretical appeal, high-order ALE methods have been increasingly employed in challenging simulation scenarios that demand both geometric flexibility and numerical accuracy. 
% Patel et al. \cite{patel_characteristic-based_2019} uses a characteristics-based approach for the spectral element method to simulate the incompressible flow during the intake stroke of an engine.
Spectral element formulations have been successfully applied to simulate complex intake stroke dynamics of internal combustion engines \citep{patel_characteristic-based_2019}, where resolving moving valves and piston motion is critical for predicting efficiency. Similarly, high-order ALE frameworks have proven instrumental in a wide range of fluid-structure interaction scenarios, providing the necessary precision to resolve the interactions between flexible structures and transient fluid flows \citep{fischer_recent_2017}. For highly dynamic regimes, such as compressible flows and explosion problems, researchers have utilized ALE methods on Voronoi meshes that support topology changes, often stabilized by a posteriori sub-cell limiters to maintain robustness without sacrificing accuracy \citep{gaburro_high_2020, gaburro_high-order_2023}. Several other studies employ high-order methods to simulate complex flow dynamics of bio-inspired flows such as flapping wings \cite{persson_numerical_2012}, fish-like movement \cite{costa_design_2020} and arterial blood flow \cite{wang_higher-order_2018}.

The Boltzmann equations, describe fluids at the macroscopic level. It has been shown that these equations can recover the Navier-Stokes equations in the low-Mach limit \cite{chapman1990mathematical}. The complex nonlinear collision term is generally replaced with relaxation models in numerical studies. Lattice-Boltzmann methods are widely used for numerical modeling of the Boltzmann equations \cite{aidun2010lattice} along with the continuous Boltzmann formulations \cite{tolke_discretization_2000, karakus_discontinuous_2019}. Numerous extensions of lattice-Boltzmann methods had been studied with ALE formulations for moving domains. It has been used for compressible flows and moving airfoils \cite{saadat2020arbitrary}, immersed moving solids \cite{meldi2013arbitrary}, incompressible flows \cite{wang2017immersed}, FSI with large deformations \cite{wu2023efficient} and many more areas of applications.

In this study, we propose an ALE form of the Galerkin-Boltzmann formulation, which gives us a moving domain solver for the nearly incompressible flow regime, implemented for GPU architectures. A nodal discontinuous Galerkin method is used for the resulting ALE form. The solver works with simplex elements in both two and three dimensions. The geometric conservation is achieved by consistent update of the geometric factors along with mesh velocity field and evaluating the spatial integrals at the same time level. A fourth order semi-analytic Runge-Kutta time integrator is used for time discretization. The remainder of this paper is structured as follows: Section \ref{sec:methodology} describes the Galerkin-Boltzmann equation system and the details of the ALE form. Sections \ref{sec:spatial_discretization} and \ref{sec:time_disc} give the details of the spatial and temporal discretization strategies. Section \ref{sec:mesh_motion_GCL} is devoted for the mesh movement strategy and the geometric conservation law. Section \ref{sec:gpu} briefly explains the GPU implementation strategies of each kernel. Lastly, in Section \ref{sec:Results}, we show numerical results validating the geometric conservation law and demonstrate the applicability of the proposed approach with concluding remarks in Section \ref{sec:conclusion}.

%% file: methodology.tex
\label{sec:methodology}

We consider the ALE form of the Galerkin-Boltzmann equations introduced in \cite{karakus_discontinuous_2019}. The methodology for the stationary form starts with the continuous Boltzmann equations describing the phase-space distribution function, $f (\mathbf{x}, \mathbf{v}, t)$ which is a function of the spatial variable $\mathbf{x}$, microscopic particle velocity $\mathbf{v}$, and time t. 
Using the Bhatnagar-Gross-Krook (BGK) collision model, the continuous Boltzmann-BGK equation becomes
\begin{equation}
    \left.\frac{\partial f }{\partial t}\right\rvert_{\mathbf{x}} + \mathbf{v}\cdot \nabla f = \frac{f^{eq}-f}{\tau},
\end{equation}
where $\tau$ is the relaxation time and $f^{eq}$ is the equilibrium phase space density. The Galerkin-Boltzmann formulation is obtained by the approach of T\"olke et al \cite{tolke_discretization_2000} where the phase space distribution function is approximated by a polynomial expansion. The unknown polynomial coefficients is obtained with a Galerkin approach, using the same polynomial for the test functions and integrating over the unbounded microscopic velocity space $\Omega_{\mathbf{v}} = (-\infty,\infty)^d$. Choosing a bi-variate Hermite polynomials lead to a constant coefficient stiffness matrix simplified as 
\begin{equation}
    \label{eq:stationary_bns}
    \left.\frac{\partial q}{\partial t}\right\rvert_{\mathbf{x}} = A_{\mathbf{x}} \cdot \nabla_{\mathbf{x}}q + \mathcal{N}(q),
\end{equation}
where $q = q(\mathbf{x},t)$ is a vector of Hermite polynomial coefficients $A_{\mathbf{x}}$ are directional coefficient matrices, and $\mathcal{N}$ is the collision operator. The detailed derivation can be found in \cite{karakus_discontinuous_2019}. A second order polynomial space approximation for the phase space yields a vector of unknown polynomial coefficients $q(\mathbf{x},t) = [q_1(x,y,t), \dots, q_n(x,y,t)]^T$, where $n=6$ for two dimensions and $n=10$ for three dimensions. The constant coefficient stiffness matrix for 2D yields 

\begin{eqnarray*}
A_x = -\sqrt{RT} \left(\begin{array}{cccccc}
0 & 1 & 0 & 0 & 0 & 0\\
1 & 0 & 0 & 0 & \sqrt{2} & 0\\
0 & 0 & 0 & 1 & 0 & 0\\
0 & 0 & 1 & 0 & 0 & 0\\
0 & \sqrt{2} & 0 & 0 & 0 & 0\\
0 & 0 & 0 & 0 & 0 & 0
\end{array}
\right),\;
A_y = -\sqrt{RT} \left(\begin{array}{cccccc}
0 & 0 & 1 & 0 & 0 & 0\\
0 & 0 & 0 & 1 & 0 & 0\\
1 & 0 & 0 & 0 & 0 & \sqrt{2} \\
0 & 1 & 0 & 0 & 0 & 0\\
0 & 0 & 0 & 0 & 0 & 0\\
0 & 0 & \sqrt{2} & 0 & 0 & 0
\end{array}
\right) 
% \]
\end{eqnarray*}
and the nonlinear collision operator is given by
\begin{eqnarray*}
\label{eq:nonlinearDef}
% \[
\mathcal{N} = -\frac{1}{\tau}\left(\begin{array}{cccccc}
0 &
0 &
0 &\
\left(q_4 - \frac{q_2q_3}{q_1}\right) &
\left(q_5 - \frac{q_2^2}{q_1\sqrt{2}}\right) &
\left(q_6 - \frac{q_3^2}{q_1\sqrt{2}}\right)
\end{array}
\right)^T.
% \]
\end{eqnarray*}
The equation system (\ref{eq:stationary_bns}) recovers the Navier-Stokes equations for low Mach number, nearly incompressible flows with kinematic viscosity $\nu=\tau RT$. The relations between macroscopic flow properties and moment of the distribution functions can be written as
\begin{eqnarray*}
\rho = q_1, \; \rho u = \sqrt{RT}q_2, \;\rho v = \sqrt{RT}q_3.
\end{eqnarray*}
The deviatoric stress tensor is given by
\begin{eqnarray*}
\sigma_{11} = -RT\left(\sqrt{2}q_5 - \frac{q_2^2}{q_1}\right), \; \sigma_{22} = -RT\left(\sqrt{2}q_6 - \frac{q_3^2}{q_1}\right)  \; \sigma_{12} = -RT\left(q_4 - \frac{q_2q_3}{q_1}\right).
\end{eqnarray*}
Lastly, the pressure is recovered through equation of state $p=\rho RT$. The details for 3D formulation can be found in \cite{ata_implicit_2026}.

To obtain the ALE form of Equation \ref{eq:stationary_bns}, the Eulerian time derivative should be replaced with ALE time derivative with respect to a fixed point in the mesh. By defining the reference frame $\boldsymbol{\xi}$ of the reference element, we can write the ALE time derivative of an arbitrary quantity $\phi$ using the Reynolds transport theorem as
\begin{equation}
    \label{eq:RTT}
    \left.\frac{\partial \phi}{\partial t}\right\rvert_{\boldsymbol{\xi}} = \left. \frac{\partial \phi}{\partial t}\right\rvert_{\mathbf{x}} + \left. \frac{\partial \mathbf{x}}{\partial t}\right\rvert_{\boldsymbol{\xi}}\cdot\frac{\partial \phi}{\partial \mathbf{x}} = \left. \frac{\partial\phi}{\partial t}\right\rvert_{\mathbf{x}} + \mathbf{v}_G\cdot\nabla_{\mathbf{x}}\phi.
\end{equation}
This theorem produces an additional transport term including $\mathbf{v}_G$ indicating the grid velocity. The mapping $\boldsymbol{f}_G(\boldsymbol{\xi},t)$ defines the motion of the domain in the physical coordinates $\mathbf{x}$ with respect to the reference frame $\boldsymbol{\xi}$. Using Equation \ref{eq:RTT}, to replace the Eulerian time derivative, the ALE form of the Galerkin-Boltzmann equation system becomes
\begin{equation}
    \label{eq:ale_bns}
    \left. \frac{\partial q}{\partial t}\right \rvert_{\xi} = A_{\mathbf{x}}\cdot\nabla_{\mathbf{x}}q + \mathbf{v}_G\cdot\nabla_{\mathbf{x}}q + \mathcal{N}(q).
\end{equation}

\begin{figure}
    \centering
    \includegraphics[width=0.6\linewidth]{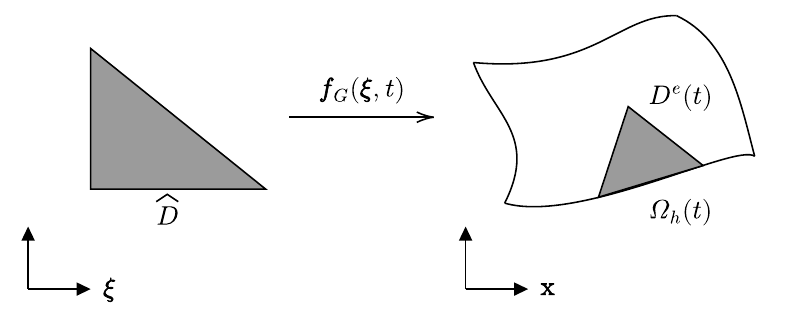}
    \caption{Illustration of the transformation of coordinate systems from the reference element $\widehat{D}$ to time dependent element $D^e(t)$ in the moving physical domain $\Omega_h$ with the mapping $\boldsymbol{f}_G(\boldsymbol{\xi},t)$, shown with two-dimensional triangle elements.}
    \label{fig:FE_mapping}
\end{figure}

%% file: dg_discretization.tex
\label{sec:spatial_discretization}

We represent the physical time dependent domain $\Omega(t)$ with computational domain $\Omega_h(t)$ composed of $K$ non-overlapping elements $D^e(t)$, where $e =1,\dots,K$ such that
\begin{equation}
    \Omega_h(t) = \bigcup_{e=1}^KD^e(t).
\end{equation}
Each element in $\Omega_h(t)$ is a straight sided simplex with an affine mapping for all $t\in[t_n, t_{n+1}]$ with the map $\boldsymbol{f}_G(\boldsymbol{\xi},t)$. The illustration of the mapping is shown in Figure \ref{fig:FE_mapping}. The mapping is time dependent, therefore the metric transformations change in every stage of the time discretization scheme. The mesh topology of $\Omega_h(t_n)$ and $\Omega_h(t_{n+1})$ is assumed to be the same such that remeshing is not considered in this work. Both $\Omega_h(t_{n})$ and $\Omega_h(t_{n+1})$ have the same number of elements and all the elements are positively oriented with respect to the reference element.

We denote the approximation to field variable $q$, on an element $D^e$ as $q^e$, and local trace values of $q^e$ on the element boundary $\partial D^e$ as $q^-$ with a corresponding neighboring trace value $q^+$. We consider a finite element space $V_n^e$ on each element to be $\mathcal{P}_N(D^e)$, the space of polynomial functions of degree $N$. As a basis of finite element spaces, we use Lagrange polynomials interpolated at the Warp \& Blend nodes \cite{warburton_explicit_2006} denoted as $\{\phi_i^e\}_{i=1}^{N_P}$. Next, starting from Equation \ref{eq:ale_bns}, we seek a local solution $q$ satisfying the following strong variational form,
\begin{equation}
    \label{eq:equation_strong_form}
    \int_{D^e}\phi \left.\frac{\partial q^e}{\partial t}\right\rvert_{\boldsymbol\xi} = \int_{D^e(t)}\phi \left( A_{\mathbf{x}}\cdot \nabla_{\mathbf{x}}q^e + \mathbf{v}_G \cdot \nabla_{\mathbf{x}}q^e \right) + \int_{\partial D^e(t)} \phi F(q^*-q^-) + \int_{D^e(t)}\phi \mathcal{N}(q^e),
\end{equation}
where $F=F_{S}+F_{ALE}$, $F_{S}=A_{\mathbf{x}}\cdot\mathbf{n}$ is the stationary flux and $F_{ALE} = \mathbf{v}_G\cdot\mathbf{n}$ is the flux coming from the ALE formulation. The numerical flux $q^*$ is defined using an upwind formulation. The $F_{S}$ is formulated as in \cite{karakus_discontinuous_2019, ata_implicit_2026} such that the operator is diagonalized as $F_{S} = \mathcal{R}\Lambda\mathcal{R}^{-1}$. The ALE upwind flux matrix will be diagonal consisting the entries of $\mathbf{v}_G$ and therefore the upwind flux can be written as,
\begin{equation}
    \label{eq:upwind_flux}
    F_{S}\;q^* = \mathcal{R}(\Lambda^+\mathcal{R}^{-1}q^- + \Lambda^-\mathcal{R}^{-1}q^+), \quad F_{ALE}\;q^* = \Lambda_{ALE}^+q^- + \Lambda_{ALE}^-q^+,
\end{equation}
where $\Lambda_{ALE}^+$ contains the positive valued mesh velocity and $\Lambda_{ALE}^-$ contains the negative valued mesh velocity. To evaluate the integral terms containing the nonlinear term $\mathcal{N}(q^e)$, we use a cubature based integration to reduce aliasing errors. We select a nodal set of $N_c$ cubature nodes on the reference element and associated weights, $w_{i}^c$ for $i = 1, \dots, N_c$. We then define a set of cubature nodes $(x_i^{e,c}, y_i^{e,c})$ on each element and map to the reference element with an interpolation operator $\mathcal{I}^e$ defined as
\begin{equation*}
    \mathcal{I}^e_{ij} = \phi_i^e(x_i^{e,c}, y_i^{e,c}).
\end{equation*}
The Jacobian of the mapping $\mathbf{x} = \boldsymbol{f}_{\mathbf{G}}(\boldsymbol{\xi}, t)$ for $\mathbf{x} = (x,y)$ and $\boldsymbol{\xi} = (r, s)$ can be defined for a time instant as
\begin{equation*}
    G^e = \begin{bmatrix}
        x_r^e & x_s^e \\ y_r^e & y_s^e
    \end{bmatrix},
\end{equation*}
and its determinant is defined as $J^e = \text{det}\;G^e$. We also define mass, surface mass and stiffness operators on each element as
\begin{equation*}
    \label{eq:operators}
    \mathcal{M}_{ij}^e = \int_{D^e(t)}\phi_j^e \phi_i^e, \quad \mathcal{M}_{ij}^{ef} = \int_{\partial D^e(t)}\phi_j^e \phi_i^e, \quad \mathcal{S}^e_{\mathbf{x}} = \int_{D^e(t)} \phi_j^e \frac{\partial \phi_i^e}{\partial \mathbf{x}}.
\end{equation*}
Using these operatores, we can write the nodal values of $q^e$ for $i = 1, \dots ,N_p$ from (\ref{eq:equation_strong_form}),
\begin{equation}
    \label{eq:discrete_1}
    \mathcal{M}_{ij}^e \frac{\partial q_j^e}{\partial t} = (A_\mathbf{x}+\mathbf{v}_G)(\mathcal{S}_\mathbf{x}^e)_{ij}q_j^e + \mathcal{M}^{ef}_{ij}(F(q^* - q^-))_j + J^e \mathcal{I}^e_{ki}w_k \mathcal{N}(\mathcal{I}^eq_j^e)
\end{equation}
for $j=1,\dots,N_p$, $k=1,\dots,N_c$ and $f=1,\dots,N_{f}$ where $N_{f}$ is the number of faces per element. By multiplying (\ref{eq:discrete_1}) by $(\mathcal{M}^e)^{-1}$, we define the differentiation, lift and projection operators as
\begin{equation*}
    \mathcal{D}_{\mathbf{x}}^e = (\mathcal{M}^e)^{-1}\mathcal{S}_{\mathbf{x}}^e, \quad\mathcal{L}^{ef} = (\mathcal{M}^e)^{-1}\mathcal{M}^{ef}, \quad \mathcal{P}^e = (\mathcal{M}^e)^{-1}(\mathcal{I}^e)^T\text{diag}(w),
\end{equation*}
respectively. Then, we can rewrite Equation \ref{eq:discrete_1} using these operators as
\begin{equation}
    \label{eq:semi_discrete_form}
    \frac{\partial q_i^e}{\partial t} = (A_\mathbf{x}+\mathbf{v}_G)(\mathcal{D}_\mathbf{x}^e)_{ij}q_j^e + \mathcal{L}^{ef}_{ij}(F(q^* - q^-))_j + J^e \mathcal{P}^e_{ik}\mathcal{N}(\mathcal{I}^eq_j^e).
\end{equation}

%% file: time_disc.tex
\label{sec:time_disc}
In our simulations, we have used fourth order semi analytic Runge-Kutta time discretization \citep{karakus_discontinuous_2019} to overcome the severe time step restriction in the limit of small relaxation time $\tau$. The nonlinear term $\mathcal{N}(q)$ in the formulation becomes stiff for small relaxation times, therefore, we write the system as
%
% \begin{equation}
%     \label{eq:time_split_stiffness}
%     \frac{dq}{dt} = \mathbf{L}(q) + \mathbf{N}(q),
% \end{equation}
%
\begin{equation}
    \frac{dq}{dt} = -\boldsymbol{\Lambda}q + \mathbf{L}(q) + \tilde{\mathbf{N}}(q),
\end{equation}
where $\boldsymbol{\Lambda} = \mathrm{diag(0, 0, 0, \frac{1}{\tau}, \frac{1}{\tau}, \frac{1}{\tau})}$ and $\Tilde{\mathbf{N}}(q) = \left( 0, 0, 0,\frac{q_2q_3}{\tau q_1}, \frac{q_2^2}{\tau q_1\sqrt{2}}, \frac{q_3^2}{\tau q_1\sqrt{2}}\right)^T$. All linear terms are collected to $\mathbf{L}$. By writing $\mathbf{F}(q) = \mathbf{L}(q) + \tilde{\mathbf{N}}(q)$ we can simplify the notation as 
\begin{equation}
    \label{eq:time_split}
    \frac{dq}{dt} = -\boldsymbol{\Lambda}q + \mathbf{F}(q).
\end{equation}
Since the nonlinear term is absent in the first three equations, these can be integrated explicitly in time with the advective time scale being the relevant stability constraint. The time-splitting approach in (\ref{eq:time_split}) separates the system such that we can apply a semi-analytic time integration method to the last three equations, while continuing to use an explicit method for the first three. The explicit scheme is designed in such a way that it aligns with the semi-analytic integration in the non-stiff limit, $1/\tau\rightarrow0$. To derive the semi-analytic formulation, we multiply (\ref{eq:time_split}) by $e^{\boldsymbol{\Lambda t}}$ and integrate in time, yielding the following expression:
\begin{equation}
    \label{eq:time_volterra}
    q(t_{n+1})= q(t_n) e^{-\boldsymbol\Lambda (t_{n+1}-t_n)} + \int_{t_n}^{t_{n+1}} e^{\boldsymbol\Lambda (\theta-t_{n+1}) }\mathbf{F}\left(q\left(\theta\right), \theta \right)d\theta.
\end{equation}
For a Runge-Kutta method, we begin by integrating (\ref{eq:time_volterra}) from \( t = t_n \) to some intermediate time level \( t = t_n + \Delta t_i \),
\begin{equation*}
q_{ni}= q_n e^{-\boldsymbol\Lambda \Delta t_i} + \int_0^{\Delta t_i} e^{\boldsymbol\Lambda(\theta-\Delta t_i) }\mathbf{F}\left(q( t_n+\theta) , t_n + \theta\right)d\theta.
\end{equation*}
Internal and final stages of a general method can be approximated by,
\begin{equation*}
\label{Eq:ERK_1}
    \begin{split}
    q_{ni}&= q_n e^{-\boldsymbol\Lambda \Delta t_i} + \Delta t\sum_{j=0}^{i-1}\tilde{a}_{ij}\mathbf{F}\left(q( t_n+\Delta t_j) , t_n + \Delta t_j\right) = q_n e^{-\boldsymbol\Lambda \Delta t_i} + \Delta t\sum_{j=0}^{i-1}\tilde{a}_{ij}\mathbf{F}_{nj}, \\
q_{n+1}&= q_n e^{-\boldsymbol\Lambda \Delta t} + \Delta t\sum_{j=0}^{s-1}\tilde{b}_{j}\mathbf{F}\left(q( t_n+\Delta t_i) , t_n + \Delta t_i\right) = q_n e^{-\boldsymbol\Lambda \Delta t} + \Delta t\sum_{j=0}^{s-1}\tilde{b}_{j}\mathbf{F}_{ni},
    \end{split}
\end{equation*}
\def\arraystretch{1.5}%  
\begin{table}
\caption{Butcher tableau for the classic RK4a, with coefficients }
    \setlength{\tabcolsep}{5pt}
      \centering
        \begin{tabular}{c| c c c c}
            0             &                &  & &\\
            $\frac{1}{2}$ & $\frac{1}{2}$  &  & &\\
            $\frac{1}{2}$ & 0  & $\frac{1}{2}$ & &\\
            1             & 0  & 0             & 1 &\\\hline
              & $\frac{1}{6}$ & $\frac{1}{3}$ &$\frac{1}{3}$ & $\frac{1}{6}$
        \end{tabular}
    \label{table:Butcher}
\end{table}
where \( s \) is the number of stages, and \( \tilde{a} \) and \( \tilde{b} \) are the semi-analytic Runge-Kutta method coefficients. Since a semi-analytic method reduces to the base Runge-Kutta method in the limit \( \frac{1}{\tau} \rightarrow 0 \), the exponential and non-exponential terms are consistent in the equation. It is assumed that the base RK method satisfies
\begin{equation}
    \sum_{j=0}^{s-1} b_j = 1, \quad \sum_{j=0}^{i-1} a_{ij} = c_i,
\end{equation}
and the semi-analytic RK scheme satisfies an analogous constraint:
\begin{equation}
    \label{eq:sark_coefficient_sum}
    \sum_{j=0}^{s-1} \tilde{b}_j = \gamma^{-1} \left( e^{\gamma} - 1 \right), \quad \sum_{j=0}^{i-1} \tilde{a}_{ij} = \frac{1}{c_i} \gamma^{-1} \left( e^{c_i \gamma} - 1 \right),
\end{equation}
for \( i = 1, \dots, s-1 \) and \( \gamma = -\frac{\Delta t}{\tau} \). Starting from the classical fourth-order Runge-Kutta method with the Butcher tableau given in Table~\ref{table:Butcher}, the coefficients of the SARK method used in this work can be presented as,
\begin{align*}
    \tilde{a}_{10} &= \gamma^{-1} \left[-1 + e^{\frac{\gamma}{2}}\right], \\
    \tilde{a}_{20} &= \gamma^{-2} \left[4 + \gamma + e^{\frac{\gamma}{2}}(\gamma-4)\right], & \tilde{b}_{0} &= \gamma^{-3} \left[ -4 -\gamma + e^{\gamma}(\gamma^2-3\gamma+4)\right], \\
    \tilde{a}_{21} &= \gamma^{-2} \left[-4 -2\gamma + 4e^{\frac{\gamma}{2}}\right], & \tilde{b}_{1} &= \gamma^{-3} \left[ 4 + 2\gamma + e^{\gamma}(2\gamma-4)\right], \\
    \tilde{a}_{30} &= \gamma^{-2} \left[2 + \gamma + e^{\gamma}(\gamma-2)\right], & \tilde{b}_{2} &= \gamma^{-3} \left[ 4 + 2\gamma + e^{\gamma}(2\gamma-4)\right], \\
    \tilde{a}_{31} &= 0, & \tilde{b}_{3} &= \gamma^{-3} \left[ -4 - 3\gamma - \gamma^2 + e^{\gamma}(4 - \gamma)\right], \\
    \tilde{a}_{32} &= \gamma^{-2} \left[-2 -2\gamma + 2e^{\gamma}\right] .
\end{align*}

%% file: mesh_motion.tex
\label{sec:mesh_motion_GCL}
To complete the ALE formulation, the grid velocity field $\mathbf{v}_G$ must be defined. Because our spatial discretization relies on an affine mapping with constant Jacobians, the physical elements are strictly required to remain straight-sided throughout the simulation. To enforce this geometric constraint for any time interval $[t_n,t_{n+1}]$, we evaluate the prescribed grid velocity at the mesh nodes and interpolate it linearly across the elements, ensuring the velocity field is globally continuous and piecewise linear $(C^0 \cup P^1)$. The nodal coordinates are advanced in time using the same Runge-Kutta integration scheme employed for the primary solver. During each Runge-Kutta stage, we evaluate the explicit, continuous velocity field $\mathbf{v}_{\mathbf{G}}$ and subsequently update the spatial coordinates of the physical nodes.

With the definition of the grid velocity, the ALE formulation of the Galerkin-Boltzmann formulation is complete. This ALE scheme should satisfy geometric conservation law, which dictates that the numerical method must exactly preserve a uniform, constant state independent of the mesh motion. To achieve this, the geometric factors are consistently updated, and all spatial integrals evaluated at the same time level
\cite{forster_geometric_2006, fehn_high-order_2021}. With consistent integration of terms, the ALE form of the Galerkin-Boltzmann formulation inherently satisfies GCL by the consistency of the spatial discretization and upwind numerical flux. To verify this, assume a uniform constant state $q(\mathbf{x},t) = q_0$ is initialized throughout the moving domain. Substituting this constant state to the right-hand side of Equation \ref{eq:equation_strong_form}, we can evaluate the terms one by one:
\begin{itemize}
    \item The volume terms including the gradient of a constant field $q_0$ is zero in a stable scheme.
    \item The consistency of the upwind flux ensures that when both the internal state $q^-$ and the neighbor state $q^+$ equal to $q_0$, the numerical flux term becomes $q^* = q_0$. Therefore the term in the surface integral $q^*-q_0$ vanishes.
    \item The collision term $\mathcal{N}(q)$ drives the system toward the equilibrium. For a uniform and constant state, the system should be perfectly at equilibrium. That yields $\mathcal{N}(q_0)=0$, vanishing the term on the right-hand side.
\end{itemize}
This yields a zero right-hand side term in Equation \ref{eq:equation_strong_form}. Applying this to Equation \ref{eq:time_split} for time integration by plugging in $q_0$,
\begin{equation*}
    \frac{d q}{\partial t} = 0 = -\mathbf{\Lambda}q_0 + \mathbf{F}(q_0),
\end{equation*}
\begin{equation*}
    \mathbf{\Lambda}q_0 = \mathbf{F}(q_0).
\end{equation*}
This means that, at the equilibrium, the explicit fluxes and nonlinear terms act as an equal and opposite force of the stiff decay $-\mathbf{\Lambda}q_0$. Plugging this analogy to our time integrator step will yield
\begin{equation*}
    \begin{aligned}
        q_{n+1} &= q_0e^{-\mathbf{\Lambda}\Delta t} + \Delta t \sum_{j=0}^{s-1} \tilde{b}_j \mathbf{F}_{ni}, \\
        q_{n+1} &= q_0e^{-\mathbf{\Lambda}\Delta t} + \Delta t \sum_{j=0}^{s-1} \tilde{b}_j \left( \mathbf{\Lambda}q_0\right).
    \end{aligned}
\end{equation*}
Factoring out the common $q_0$ yields
\begin{equation*}
    q_{n+1} = q_0 \left( e^{-\mathbf{\Lambda}\Delta t} + \Delta t \mathbf{\Lambda} \sum_{j=0}^{s-1}\tilde{b}_j \right).
\end{equation*}
Satisfying the constraint in Equation \ref{eq:sark_coefficient_sum}, with $\gamma = -\mathbf{\Lambda}\Delta t$, 
\begin{equation*}
    \begin{aligned}
        q_{n+1} &= q_0 \left( e^{-\mathbf{\Lambda}\Delta t} + \Delta t \mathbf{\Lambda} (\gamma^{-1}\left( e^{\gamma} -1\right) ) \right), \\
        q_{n+1} &= q_0 \left( e^{-\mathbf{\Lambda}\Delta t} + \Delta t \mathbf{\Lambda} (-\mathbf{\Lambda}\Delta t)^{-1}\left( e^{-\mathbf{\Lambda}\Delta t} -1\right) \right).
    \end{aligned}
\end{equation*}
The expression in the parentheses becomes unity and this semi-analytic Runge-Kutta scheme yields $q_{n+1} = q_0$ in every timestep, satisfying the geometric conservation law. The numerical tests in Section \ref{subsec:GCL_results} shows that the free stream state can be preserved in the ALE form of the Galerkin-Boltzmann formulation.

%% file: GPU_implementation.tex
\label{sec:gpu}
In this section, we give a brief information about the implementation for obtaining the numerical results. The formulation described above is implemented into the \texttt{libParanumal} kernel library \cite{ChalmersKarakusAustinSwirydowiczWarburton2020}, using C++ and OCCA kernel language \cite{medina2014occa}. OCCA is a vendor neutral framework supporting multiple backends, offering flexibility for building portable device kernels. 

The solution process consists of six fundamental compute kernels. i) interpolation of grid velocities on solution nodes, ii) recalculating the geometric factors, iii) evaluation of volume integrals, iv) evaluation of surface integrals, v) cubature integration of relaxation term, vi) timestep update. We refer to these kernels as; velocity interpolation kernel, geometric factor kernel, volume kernel, surface kernel, cubature kernel and update kernel respectively.

\begin{itemize}
    \item \textbf{Velocity Interpolation Kernel}: In the numerical implementations, the grid velocity field is explicitly defined on the vertices of the elements. For the affine mapping, we interpolate the vertex velocities to the interpolation nodes and obtain the grid velocity in the space of $P^1$. The vertex velocity field is first loaded from global memory to shared memory. An interpolation array of size $N_p\times N_{vertices}$ is re-used within the kernel by taking advantage of L1 or L2 caches. Every thread calculates the inner product of a row of the interpolation matrix and the vertex velocity vector stored in the shared memory. 

    \item \textbf{Geometric Factor Kernel}: After the mesh moves, the geometric factors, including the elemental Jacobian, spatial derivative matrices, and interfacial face normals, must be re-evaluated. Since the deformation preserves the affine simplicial elements, these geometric factors remain constant throughout an element and determined by the coordinates of linear vertices. Therefore, rather than assigning a thread block to a single element, the global array is segmented into discrete tiles of block size 256, assigning exactly one thread to compute the geometry for one entire element. Once the vertices are loaded, all geometric computations are localized entirely within registers before the updated metrics are written back to global memory.

    \item \textbf{Volume Kernel}: The ALE-modified volume kernel evaluates the semi-discrete volume integral terms in (\ref{eq:semi_discrete_form}). Similar to the static mesh formulation, the kernel initially loads the nodal solution fields from global device memory into shared memory arrays of size $N_p$. The differentiation matrices are accessed directly from global memory, exploiting L1 or L2 hardware caches to rapidly broadcast the static reference element operators. Spatial derivatives are evaluated at each node via inner products between the cached matrix rows and the shared solution vectors. Under the restriction of affine mesh transformations, the geometric factors and the Jacobian of local to global mapping remain uniform across the element. Therefore, a single set of geometric metrics is loaded per element and held in register memory. To evaluate the ALE transport operator, the nodal mesh velocities are fetched from global memory utilizing fully coalesced read access patterns. By applying the mesh velocities directly to the physical spatial derivatives already available in the thread registers, the volume integral is calculated.

    \item \textbf{Surface Kernel}: The surface kernel computes the interfacial flux contributions of the semi-discrete formulation (\ref{eq:semi_discrete_form}). Similar to the volume kernel, this kernel loads the trace data of the element and all its neighbor node data, including the mesh velocities, into the register memory. Since the mesh velocity field is $C^0$ continuous, it is the same for the trace and neighbor data. These data along with the re-calculated surface geometric factors are then used in numerical flux routines. The resulting fluxes are then scaled by the geometric data and stored in a shared memory array size $N_f\times N_{fp}$, where $N_f$ is the number of faces and $N_{fp}$ is the number of nodes along each element face. The computed flux arrays are then lifted to the interpolation nodes with the matrix-vector multiplication similar to the differentiation in volume kernel. 

    \item \textbf{Cubature Kernel}: The nonlinear relaxation term in the semi-discrete form (\ref{eq:semi_discrete_form}) are evaluated with an appropriately high-order cubature rule. The field variables are first copied to shared memory using $N_p$ threads. Then, these values are interpolated to the cubature integration points using the interpolation matrix $\mathcal{I}$ with $N_c$ threads, $N_c$ being the number of cubature points. The nonlinear term is computed on cubature nodes and stored in shared memory arrays. Finally, the results are projected back with the projection operator $\mathcal{P}$ to the interpolation nodes.

    \item \textbf{Update Kernel}: This kernel performs the time integration updates with global vector operations using the right hand side vectors. The Runge-Kutta coefficients are stored in register memory for quick access for the update procedures with $N_p$ threads.
\end{itemize}

%% file: results.tex
\label{sec:Results}

In this section, we present the numerical results to show the accuracy and scalability of the ALE form of the Galerkin-Boltzmann equation. We select different test cases showing different aspects of the proposed formulation. A free stream preservation test showing the geometric conservation property of the solver is presented first. Then, a three-dimensional Taylor-Green vortex solution is presented to show the performance and scalabilty of the solver in large problems. Lastly, we study a plunging airfoil test case for moving aerodynamics problems. In all of the test cases, the grid velocity field $\mathbf{v}_G$ is explicitly defined in the whole domain. The solutions are obtained in Marenostrum 5 Accelerated partition of Barcelona Supercomputing center with NVIDIA H100 GPUs.

\subsection{Geometric Conservation Law - Free stream preservation tests}
\label{subsec:GCL_results}
\begin{figure*}
    \centering
    \begin{subfigure}[b]{0.45\textwidth}
        \centering
        \includegraphics[width=\textwidth]{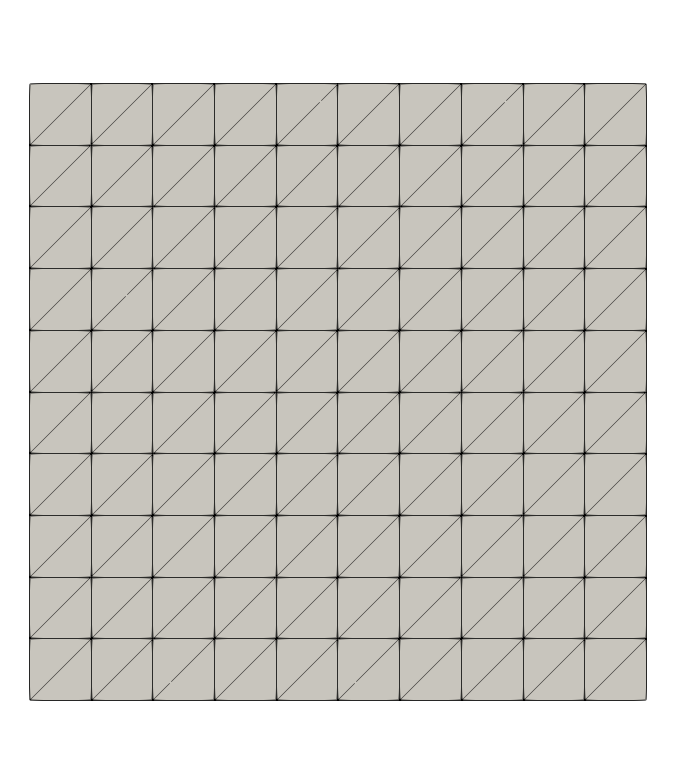}
        \caption{Node positions at $t=0$}
    \end{subfigure}%
    ~ 
    \begin{subfigure}[b]{0.45\textwidth}
        \centering
        \includegraphics[width=\textwidth]{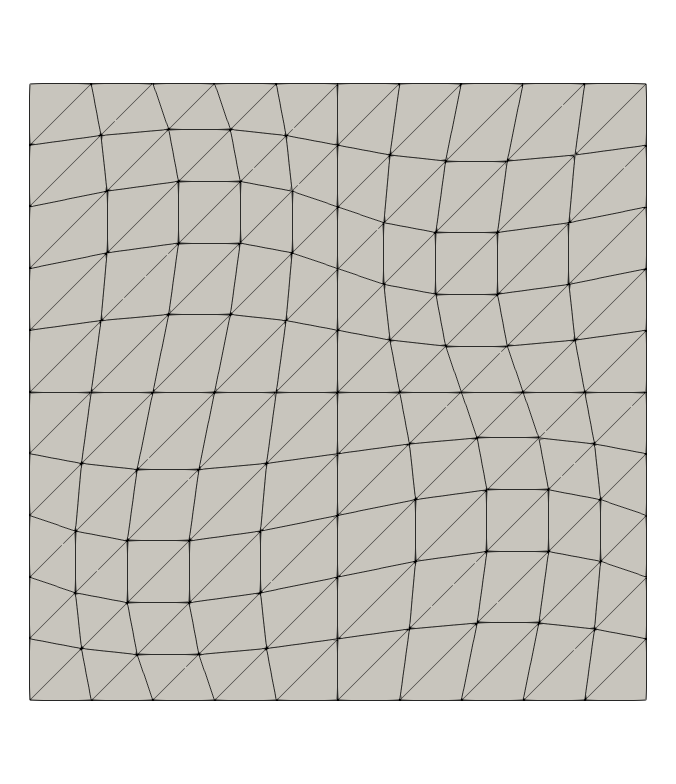}
        \caption{Node positions at $t=20$}
    \end{subfigure}
    \caption{The initial and final positions of the free stream preservation test case.}
    \label{fig:GCL_test}
\end{figure*}
We conduct the free stream preservation test to observe if the ALE form of the Galerkin-Boltzmann formulation fulfills the geometric conservation law. The solver should be able to preserve the free stream flow conditions and the flow field should not be disturbed by a moving mesh. We adapt the test case from \cite{nguyen_arbitrary_2010} where the grid motion is defined as:
\begin{equation}
    \label{eq:gcl_exact_field}
    \begin{aligned}
        x(t) &= x_0 + X_0\sin{(n_t2\pi t /t_0)}\sin{(n_x2\pi x/L_x)}\sin{(n_y2\pi y/L_y)} \\
        y(t) &= y_0 + Y_0\sin{(n_t2\pi t /t_0)}\sin{(n_x2\pi x/L_x)}\sin{(n_y2\pi y/L_y)},
    \end{aligned}
\end{equation}
where $(x,y)$ are the grid coordinates in two-dimensional space and $(x_0,y_0)$ are the grid coordinates before the movement. Here, $n_x,\;n_y$ and $n_t$ are the periods in space and time taken as $n_x=n_y=n_t=1$. The amplitude of the deformation is given as $X_0=Y_0=0.5$. $L_{x,y}$ are the domain size in each direction and they both are $L_x=L_y=20$. For the free stream case we equate the density $\rho=1$, and velocities in $x$ and $y$ directions $u=1,\;v=1$ respectively for the initial condition. The boundaries specified as periodic boundary condition to not disturb the field. The simulation parameters are set such that $Ma=0.1$ and $Re=100$. The simulation is run for over a time interval $0\leq t \leq T=20$. The grid positions at $t=0$ and $t=T=20$ can be seen in Figure \ref{fig:GCL_test}. The positions of the elements are updated by solving $d\boldsymbol{x}/dt = \mathbf{v}_G$ where $\mathbf{v}_G$ is found with the time derivative of Equation \ref{eq:gcl_exact_field}. We calculate the $L_2$ norm of the error of the momentum in both directions, which are the conserved quantities $\rho u$ and $\rho v$. The numerical results for refining number of elements is presented in Table \ref{tab:GCL_Error}. The results show that the error is at order of $10^{-15}$ showing the geometric conservation law is satisfied under these settings.

\begin{table}
    \centering
    \caption{$L_2$ norm of error of the conserved variables versus free flow condition at $t=10$ for different number of elements $K$}
    \begin{tabular}{c c c c c c c}
    \hline \hline 
       $K$           &  & $20^2$  & &  $40^2$  & & $80^2$ \\ \hline
       $||\rho u||$  &  & $1.11\times10^{-15}$ & & $2.22\times10^{-15}$ & & $4.44\times10^{-15}$ \\
       $||\rho v||$  &  & $1.11\times10^{-15}$ & & $2.22\times10^{-15}$ & & $4.44\times10^{-15}$ \\
    \hline \hline
    \end{tabular}
    \label{tab:GCL_Error}
\end{table}

\subsection{3D Taylor-Green Vortex}
We study the behavior of the ALE form of the Galerkin-Boltzmann formulation in transitional and turbulent flows using the three-dimensional Taylor-Green vortex problem. This is a benchmark problem showing the dissipation of energy and decay of isotropic turbulence. The flow starts with the initial state,
\begin{align}
    u(x,y,z,t_0)&=U_{0}sin(x/L)cos(y/L)cos(z/L), \\  
    v(x,y,z,t_0)&=-U_{0}sin(y/L)cos(x/L)cos(z/L), \\
    w(x,y,z,t_0)&=0.
\end{align}
The solution domain starts with $\Omega_0 = [-\pi, \pi]^3$ that deforms over time according to the deformation in Equation \ref{eq:TGV_deformation}, adapted from \cite{fehn_high-order_2021}.
\begin{equation}
    \label{eq:TGV_deformation}
    \begin{aligned}
        x(t) &= X + A\sin{\left(2\pi \frac{t}{T_G}\right) \sin{\left( 2\pi \frac{Y +L/2}{L}\right)}} \sin{\left( 2\pi \frac{Z + L/2}{L}\right)} \\
        y(t) &= Y + A\sin{\left(2\pi \frac{t}{T_G}\right) \sin{\left( 2\pi \frac{X +L/2}{L}\right)}} \sin{\left( 2\pi \frac{Z + L/2}{L}\right)} \\
        z(t) &= Z + A\sin{\left(2\pi \frac{t}{T_G}\right) \sin{\left( 2\pi \frac{X +L/2}{L}\right)}} \sin{\left( 2\pi \frac{Y + L/2}{L}\right)},
    \end{aligned}
\end{equation}
where $X, \;Y,\;Z$ are the initial positions in three different directions, $L=2\pi$ is the length of the initial cube and $A=\pi/6$ is the amplitude. The simulated time is $0 \le t \le 20$ and the $T_G=20$. The Reynolds number is selected as $Re=1600$ and Mach number is set to $Ma=0.1$ to be in nearly incompressible regime. The computational domain is decomposed to 32 cubes in all three dimensions with each cube is split to 6 tetrahedron which results in a total of $K=196608$ elements. The results are obtained with polynomial order $N=3$ and compared with stationary version of Galerkin-Boltzmann solver \cite{karakus_discontinuous_2019} and results obtained using a spectral element method with $521^3$ elements \cite{van2011comparison}.
\begin{figure}
        \centering
        \begin{subfigure}[b]{0.48\textwidth}
         \includegraphics[width=\textwidth]{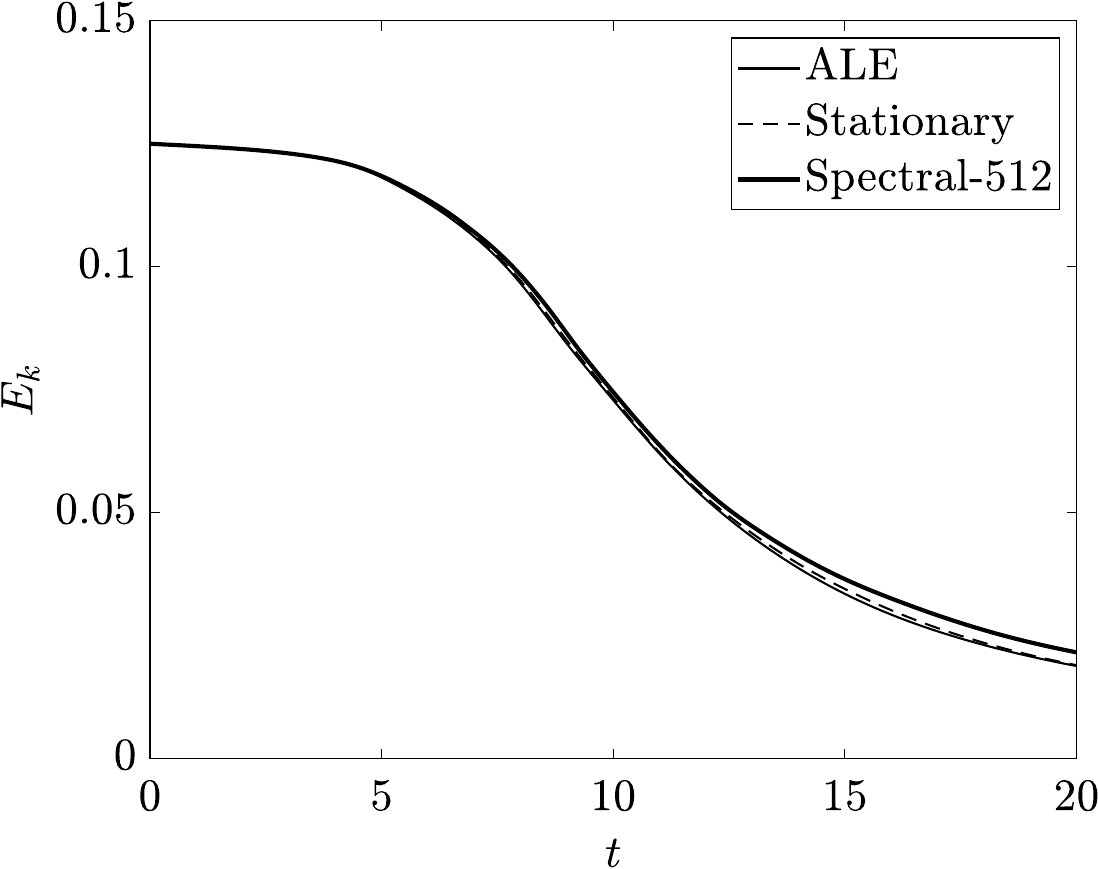}
        \caption{Evolution of the kinetic energy}
    \label{fig:TGV_KE}    
        \end{subfigure}
    ~
     \begin{subfigure}[b]{0.48\textwidth}
         \includegraphics[width=\textwidth]{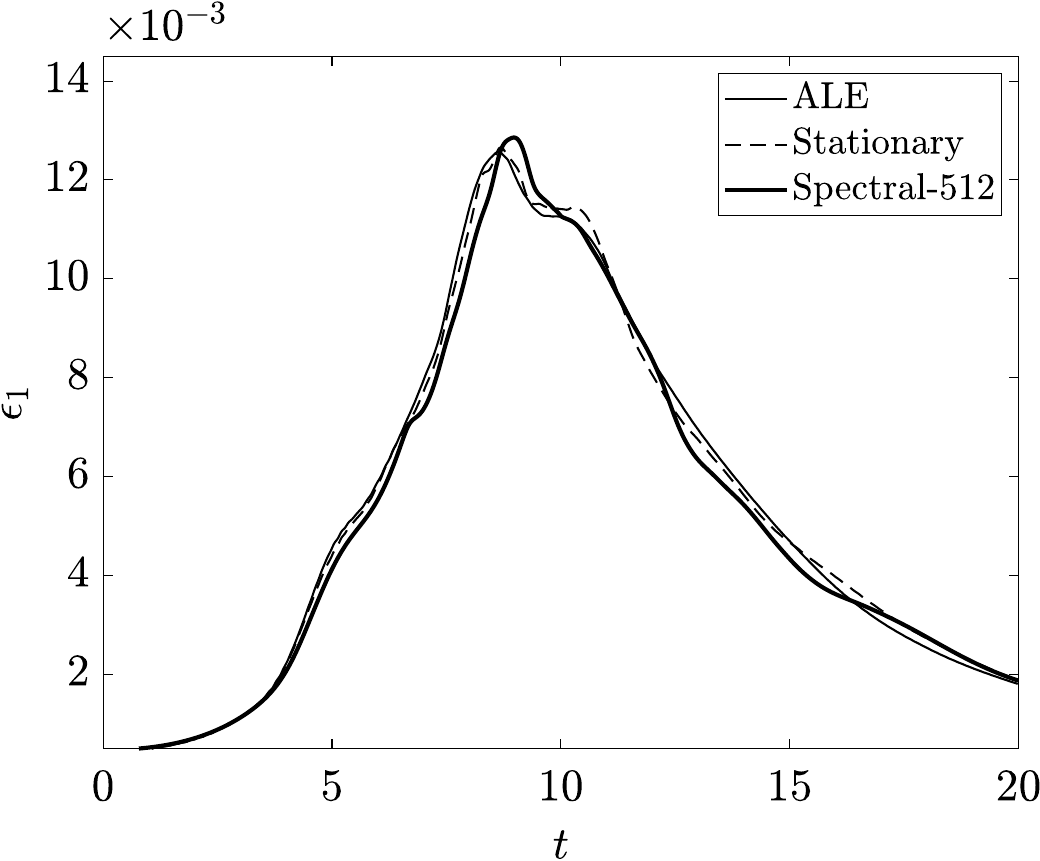}
        \caption{Kinetic energy dissipation rate}
    \label{fig:epsilon1_TGV}    
        \end{subfigure}
    ~
    \caption{Kinetic energy and its dissipation rate for the moving 3D TGV.}
    \label{fig:KE_and_epsilon}
\end{figure}
We analyze the problem accuracy by comparing the evolution of the kinetic energy calculated as
\begin{equation*}
    E_K = \frac{1}{\rho_{\infty}\Omega} \int_{\Omega}\frac{1}{2}\rho \mathbf{u}\cdot \mathbf{u}\, d\Omega
\end{equation*}
where the reference density is $\rho_{\infty}=1$. The time evolution of the kinetic energy between $0 \le t \le 20$ is shown in Figure \ref{fig:TGV_KE}. The ALE formulation can follow the same trend as the stationary solver and together they slightly dissipate in final time compared with the spectral solution since tetrahedral elements introduce more dissipation than the spectral solvers \cite{pereira_spectral_2020}.

The kinetic energy dissipation rate $\epsilon_1 = -dE_k/dt$ is also compared with the stationary form and the reference spectral solution in Figure \ref{fig:epsilon1_TGV}. Comparison with the stationary form show that the peak is achieved at a similar time, showing the ALE formulation does not introduce significant dissipation. The peaks are slightly earlier than the reference solution due to a lower resolution in our results. 

We also tested the performance of our ALE solvers on multi-GPU systems in Marenostrum 5 to observe the performance of our ALE solver. We measured the average wall time per timestep, $t_{step}$, using 1000 time steps with the semi analytic Runge-Kutta timestepper. The scaling results are presented in Table \ref{tab:weak_scaling_tgv}, where the total degrees of freedom (DoF) are given as an approximate number. We started with one node and performed our analysis up to 16 nodes where we placed 82944 elements per GPU. The solver has a good weak scaling performance as the wall time per time step stays very similar with different number of nodes. The 16 node solution surpassed a total of one billion degrees of freedom showing the applicability of the ALE solver in large scale simulations.

\begin{table}
    \centering
    \caption{Weak-scaling study of the 3D moving Taylor-Green vortex}
    \label{tab:weak_scaling_tgv}
    \begin{tabular}{c c c c c c}
    \hline \hline
    Nodes & \# of GPUs & $K$ & Total DoF &$t_{step}$ (s) & eff \% \\
    \hline
    1  & 4  & 331776  & 66M   & 1.418$\times 10^{-2}$ & 100 \\
    2  & 8  & 663552  & 132M  & 1.632$\times 10^{-2}$ & 86.8 \\
    4  & 16 & 1327104 & 265M  & 1.578$\times 10^{-2}$ & 89.8 \\
    8  & 32 & 2654208 & 530M  & 1.556$\times 10^{-2}$ & 91.1 \\
    16 & 64 & 5308416 & 1B & 1.588$\times 10^{-2}$ & 89.3 \\
    \hline \hline 
    \end{tabular}
\end{table}

\subsection{Flow Over a Plunging Airfoil}
In this section, we considered nearly incompressible flow over plunging NACA0012 airfoil for varying Strouhal Numbers. Variation in plunging frequencies results in vortex dominated wake flows with different characteristics. Besides increasing the Strouhal number results in thrust generation after a point which is known as Knoller–Betz effect. 
% In this context we investigated three different cases where negative drag, zero drag and thrust production occurs.
The results are compared with the results from both experimental \cite{jones1998experimental} and numerical studies \cite{liang2011high,saadat2020arbitrary}. Time dependent and averaged lift and drag values, vorticity visualizations are shared and compared. The sinusoidal plunging motion of the airfoil is given by
\begin{align}
y(t)&=Y-h_0sin(\omega t), \\
x(t)&=X,    
\end{align}
where $X$, $Y$ are initial positions and $h$, $\omega$ are amplitude and frequency of the plunging motion, respectively. The Strouhal number is described as
\[St=\frac{h\omega}{u_\infty}\]
where $u_\infty$ is the non-dimensional free-stream velocity. The Reynolds number is set to $Re=1850$ and the Mach number is $Ma=0.1$. Three different cases considered in this problem are Case 1: $St=0.29$ with $h_0=0.08$, Case 2: $St=0.46$ with $h_0=0.08$, and Case 3: $St=0.60$ with $h_0=0.20$. These configurations result in drag production, zero drag, and thrust production, respectively according to simulations performed using a panel code and experimental results \cite{jones1998experimental}.

\begin{table}
    \centering
    \begin{tabular}{c c c c c c c}
    \hline \hline      
        & \multicolumn{2}{c}{Case 1} 
        & \multicolumn{2}{c}{Case 2} 
        & \multicolumn{2}{c}{Case 3} \\
        \cline{2-3} \cline{4-5} \cline{6-7}
        & $St=0.29$ & $h_0=0.08$ 
        & $St=0.46$ & $h_0=0.08$ 
        & $St=0.60$ & $h_0=0.20$ \\
    \hline
        $c_D$ & \multicolumn{2}{c}{0.043} & \multicolumn{2}{c}{-0.029} & \multicolumn{2}{c}{-0.077} \\
        $c_L$ & \multicolumn{2}{c}{-0.006} & \multicolumn{2}{c}{-0.016} & \multicolumn{2}{c}{-0.013} \\        
    \hline \hline
    \end{tabular}
    \caption{Average drag and lift coefficients of the plunging airfoil test cases}
    \label{tab:drag_and_lift}
\end{table}

\begin{figure}
        \centering
        \begin{subfigure}[b]{0.31\textwidth}
         \includegraphics[width=\textwidth]{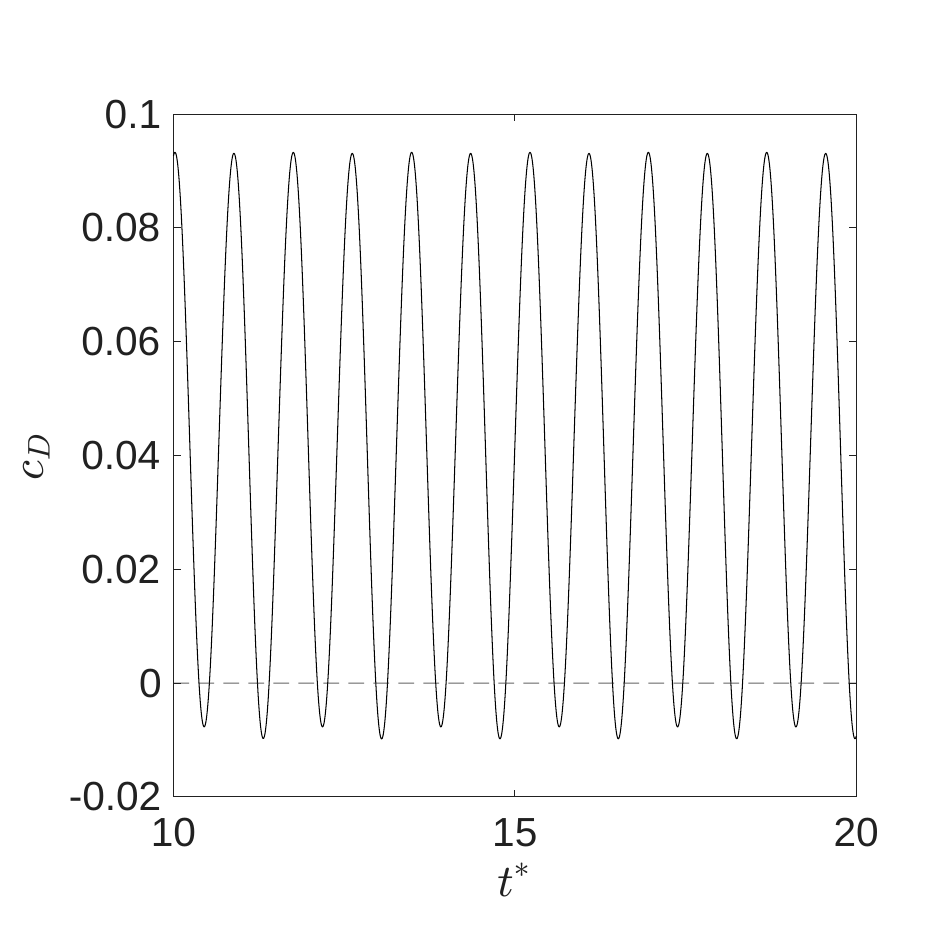}
        \caption{Case 1}
    \label{fig:Drag_1}    
        \end{subfigure}
    ~
     \begin{subfigure}[b]{0.31\textwidth}
         \includegraphics[width=\textwidth]{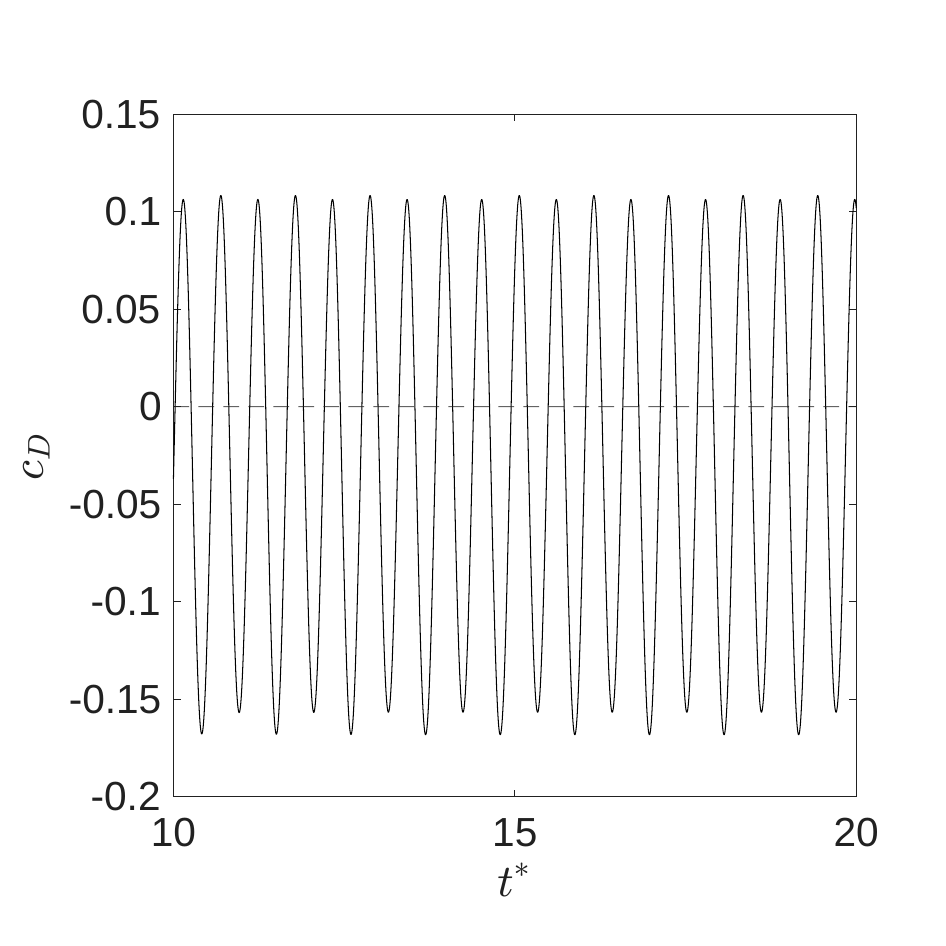}
        \caption{Case 2}
    \label{fig:Drag_2}    
        \end{subfigure}
    ~
    \begin{subfigure}[b]{0.31\textwidth}
         \includegraphics[width=\textwidth]{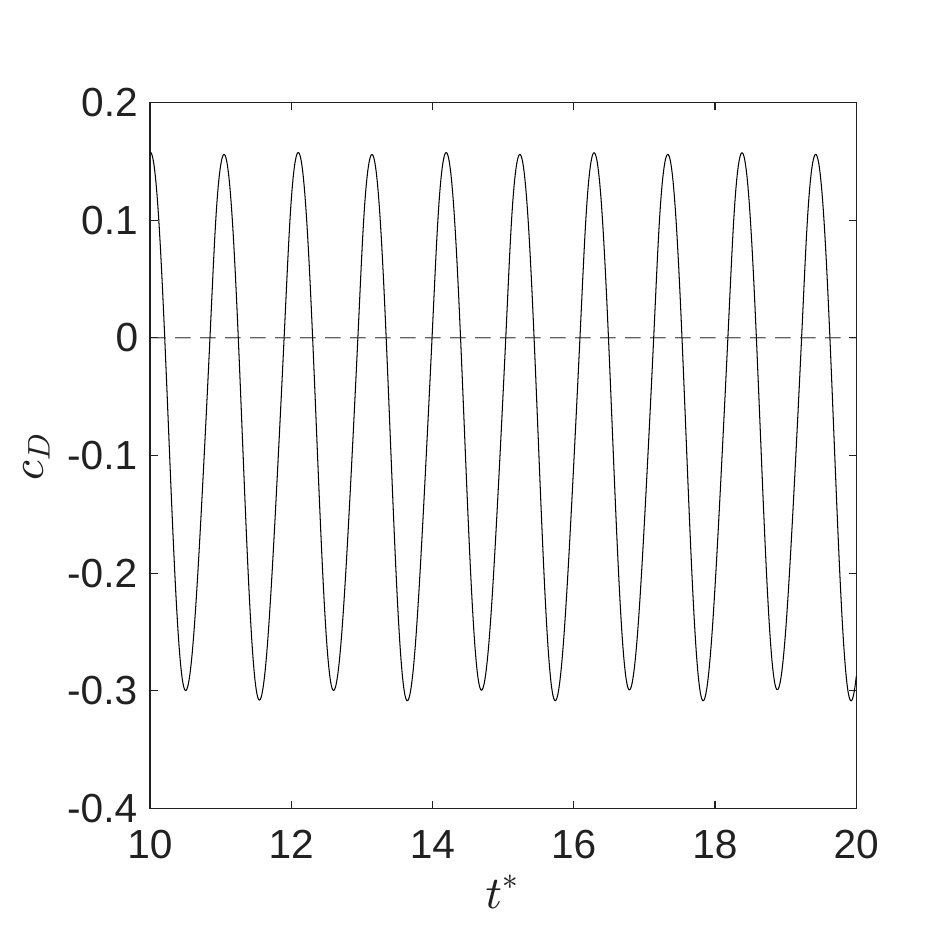}
        \caption{Case 3}
    \label{fig:Drag_3}    
        \end{subfigure}
    ~
    \caption{Time evolution of drag coefficient of different cases.}
    \label{fig:plunging_drag}
\end{figure}

\begin{figure}
        \centering
        \begin{subfigure}[b]{0.48\textwidth}
         \includegraphics[width=\textwidth]{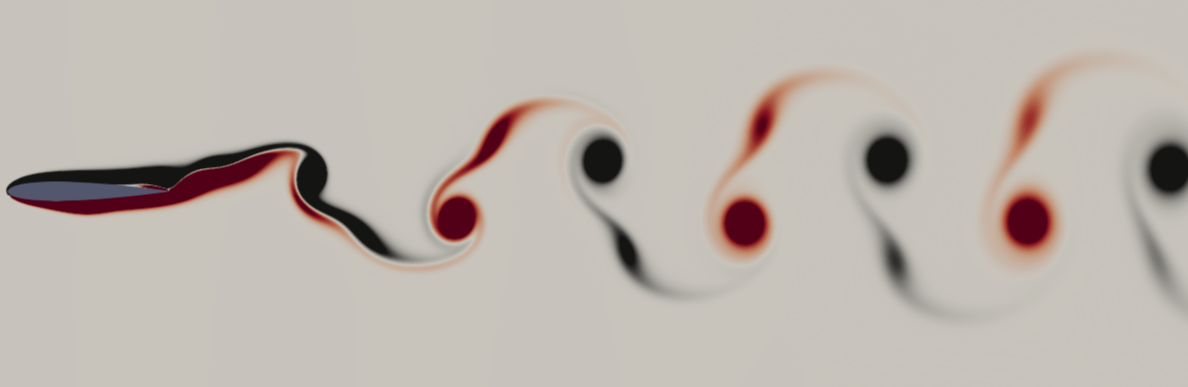}
        \caption{Case 1}
    \label{fig:Vortex_1}    
        \end{subfigure}
        \begin{subfigure}[b]{0.48\textwidth}
         \includegraphics[width=\textwidth]{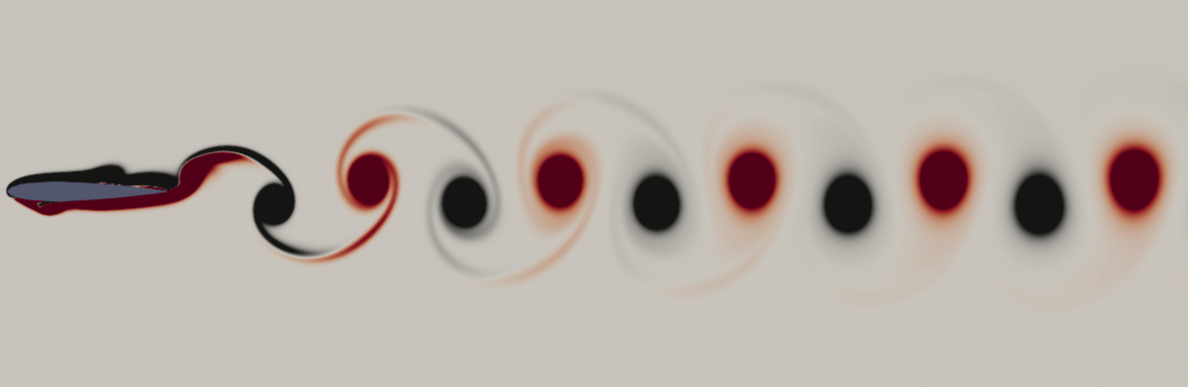}
        \caption{Case 2}
    \label{fig:Vortex_2}    
        \end{subfigure}
        \begin{subfigure}[b]{0.48\textwidth}
         \includegraphics[width=\textwidth]{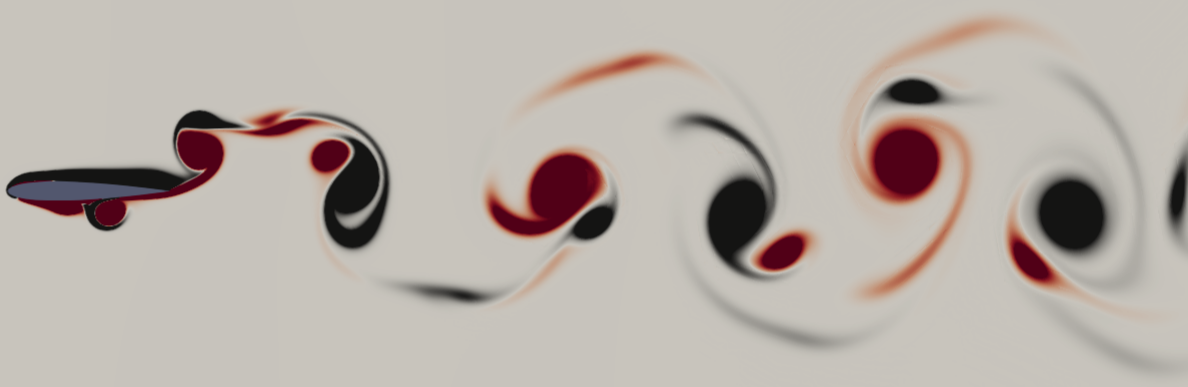}
        \caption{Case 3}
    \label{fig:Vortex_3}    
        \end{subfigure}
        \caption{Vortex structures behind the plunging airfoil.}
    \label{fig:vorticity_plunging_airfoil}   
\end{figure}

The time evolution of the drag coefficients for all cases are presented in Figure \ref{fig:plunging_drag}. The drag coefficient is calculated as $c_D = F_D/(0.5\rho_{\infty}u_{\infty}^2c)$ where $F_D$ is the total drag force and $c=1$ is the chord length. As it can be seen in Figure \ref{fig:plunging_drag}, the flow produces net drag for the Case 1 with an average drag coefficient of $c_D=0.043$. The vortex structures behind the airfoil are presented in Figure \ref{fig:vorticity_plunging_airfoil}. The wake pattern of Case 1 shows the characteristic von K\'arm\'an vortex street indicating the drag production. As for the Case 2, the experimental studies indicate zero drag production. Although the vortex structure in Figure \ref{fig:vorticity_plunging_airfoil} is similar to the experimental study with almost symmetric vortex structures around the center of the domain \cite{jones1998experimental}, the drag coefficient oscillates around a negative value similar to the numerical studies in \cite{liang2011high, saadat2020arbitrary}. Our simulation reveals an average drag coefficient of $c_D=-0.029$ indicating a small amount of thrust is generated in the zero drag case. The thrust production in Case 3 is validated by the average negative drag coefficient $c_D = -0.077$. Also, the vortex positions are in agreement with the thrust production stated in \cite{jones1998experimental}. All three cases produces almost zero average lift coefficient. These values, along with the average drag coefficients are tabulated in Table \ref{tab:drag_and_lift}.

\subsection{Flow Over a Carangiform Fish}
\begin{figure}[hbtp!]
    \centering
    \includegraphics[width=0.8\linewidth]{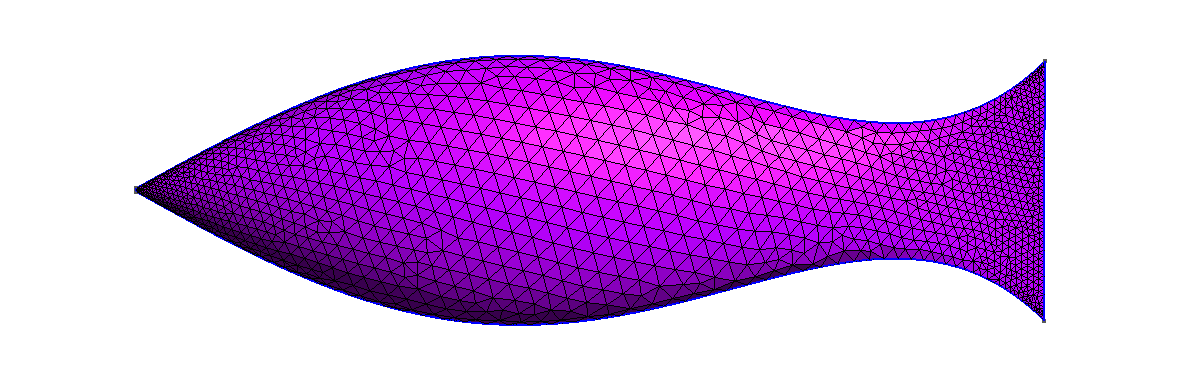}
    \caption{Fish geometry and the surface mesh}
    \label{fig:fish_surface_mesh}
\end{figure}

In this section, a carangiform fish and its swimming motion are modeled in the nearly incompressible flow medium. The geometry is adapted from \cite{valdivia2007design}, by defining elliptical profiles through the fish body. The half-height $R(x)$ in the $y$ direction and half-width $r(x)$ in $z$ direction measured from the center is defined as 
\begin{equation}
    \begin{aligned}
        R(x) &= 0.14L \sin{(2\pi x/1.6L) + 0.0008L(e^{2\pi x/1.1L} - 1)} \\
        r(x) &= 0.045L\sin{(2\pi x / 1.25L)} + 0.06L \sin{(2\pi x /3.14L)},
    \end{aligned}
\end{equation}
where $L$ is the full-length of the fish body and $x$ is the axial coordinate along the fish body starting from the nose. The body and the surface mesh is visualized with $L=1$ in Figure \ref{fig:fish_surface_mesh}. The swimming motion of a carangiform fish is defined as the following displacement of the centerline of the fish body:
\begin{equation}
    \label{eq:fish_motion}
    \Delta z(x,t) = A(x) \sin{(2\pi x /\lambda - 2\pi f t)},
\end{equation}
where $f$ is the tail beat frequency and $\lambda$ is the wavelength. $A(x)$ is the amplitude function given as
\begin{equation}
    \label{eq:fish_amplitude_function}
    A(x)/L = a_0 + a_1(x/L) + a_2(x/L)^2.
\end{equation}
Seo et al. \cite{seo_scaling_2026} give a set of parameters to define the carangiform fish movement with $a_0=0.02$, $a_1=-0.08$, $a_2=0.16$. Since the wavelength $\lambda$ is close to the body length $L$ \cite{videler1984fast,seo_scaling_2026}, it is set to $\lambda=L$. 

The mesh motion is prescribed in a purely explicit ALE form by updating only the $z$ component of each mesh vertex, while keeping the streamwise and spanwise coordinates unchanged. We first define a clamped axial coordinate with the initial mesh positions $(x_0, y_0,z_0)$:
\[
\tilde{x}=\min\bigl(1,\max(0,x_0)\bigr),
\]
and construct an effective distance to a rigid core of radius $R_{\mathrm{rigid}}$ via
\[
d_{\mathrm{eff}}=\max\!\left(\sqrt{(x_0-\tilde{x})^2+y_0^2+z_0^2}-R_{\mathrm{rigid}},\,0\right).
\]
The spatial blending function $B$ is then defined as
\[
B(d_{\mathrm{eff}})=\exp\!\bigl(-\beta_{\mathrm{decay}}\,d_{\mathrm{eff}}^2\bigr),
\]
and the local oscillation amplitude is
\[
A(\tilde{x})=a_0+a_1\tilde{x}+a_2\tilde{x}^2,\qquad
\phi(x_0,t)=k\,x_0-\omega t.
\]
This blending function transmits the boundary movement to the mesh. To avoid an impulsive start, a cosine ramp $S(t)$ is applied:
\[
S(t)=
\begin{cases}
\frac12\!\left(1-\cos\!\left(\pi t/T_{\mathrm{ramp}}\right)\right), & t<T_{\mathrm{ramp}},\\[4pt]
1, & t\ge T_{\mathrm{ramp}},
\end{cases}
\quad
\dot S(t)=
\begin{cases}
\frac{\pi}{2T_{\mathrm{ramp}}}\sin\!\left(\pi t/T_{\mathrm{ramp}}\right), & t<T_{\mathrm{ramp}},\\[4pt]
0, & t\ge T_{\mathrm{ramp}}.
\end{cases}
\]
Hence the imposed displacement and mesh-velocity field are
\[
h(x_0,y_0,z_0,t)=A(\tilde{x})\sin\phi\,B\,R,
\]
\[
w_z(x_0,y_0,z_0,t)=\frac{\partial h}{\partial t}
=\Bigl[-A(\tilde{x})\,\omega\cos\phi\,B\Bigr]S
+\Bigl[A(\tilde{x})\sin\phi\,B\Bigr]\dot S,
\]
with $w_x=w_y=0$, which yields a smooth, spatially localized fish-induced deformation with a controlled temporal ramp-up. For the simulations, $T_{\mathrm{ramp}}$ is selected as one tail-beat period, the rigid core is selected as $R_{\mathrm{rigid}}=0.5$  with the decay rate $\beta_{\mathrm{decay}}=10.0$.

The simulations are performed in various Reynolds number regimes with a corresponding Strouhal number defined as $St=fA_f/U$, where $A_f$ is the peak to peak tail beat amplitude which becomes $A_f/L=0.2$ with the selected motion parameters. The investigation of the model is through the force coefficients in the flow direction with a set of flow conditions adapted from \cite{seo_scaling_2026}. The coefficients are calculated as 
\[
C_p = \frac{F_p}{\frac12 \rho u^2A}, \quad C_v = \frac{F_v}{\frac12 \rho u^2A},
\]
where the subscript $p$ represents the pressure force contribution and $v$ represents the viscous force contributions. The frontal area of the fish, $A$, is approximately 0.039$L^2$ with the geometry used in this study. The contributions of pressure and viscous forces are separated for the body and the caudal fin of the fish. The force coefficients for various flow conditions are tabulated in Table \ref{tab:force_fish}. These solutions are obtained with $K=324165$ elements and polynomial order $N=4$. Perfectly matched layers are used to damp out the oscillations from the boundaries. The negative values in Table \ref{tab:force_fish} indicates thrust and positive values indicate drag. In the body of the fish, viscous drag forces dominate the pressure thrust. On the contrary, the fin produces thrust from pressure forces, dominating the viscous drag on the fin, physically consistent with the results from the literature \cite{huang_effect_2024,seo_scaling_2026}.
\begin{table}
    \centering
    \caption{Time averaged force coefficients of the carangiform fish for different flow conditions. The individual contributions of the body and the fin are presented separately.}
    \label{tab:force_fish}
    \begin{tabular}{c c c c c c}
    \hline \hline
    $Re$ & $St$ & $C_{p,\mathrm{body}}$ & $C_{v,\mathrm{body}}$ & $C_{p,\mathrm{fin}}$ & $C_{v,\mathrm{fin}}$ \\
    \hline
    20    & 0.56 & -0.35 & 1.14  & -0.78 & 0.39 \\
    2400   & 0.42 & -0.10 & 0.51  & -0.35 & 0.14  \\
    5800   & 0.34 & -0.02 & 0.30  & -0.19 & 0.07   \\
    17000  & 0.29 &  0.004 & 0.17  & -0.12 & 0.038  \\
    36000  & 0.28 & -0.009 & 0.12  & -0.094 & 0.027  \\
    % 720    & 0.56 & -0.35 & 1.14  & -0.78 & 0.39 \\
    % 2400   & 0.42 & -0.10 & 0.51  & -0.35 & 0.14  \\
    % 5800   & 0.34 & -0.02 & 0.30  & -0.19 & 0.07   \\
    % 17000  & 0.29 &  0.004 & 0.17  & -0.12 & 0.038  \\
    % 36000  & 0.28 & -0.009 & 0.12  & -0.094 & 0.027  \\
    \hline \hline 
    \end{tabular}
\end{table}

\begin{figure}
        \centering
        \begin{subfigure}[b]{0.48\textwidth}
         \includegraphics[width=\textwidth]{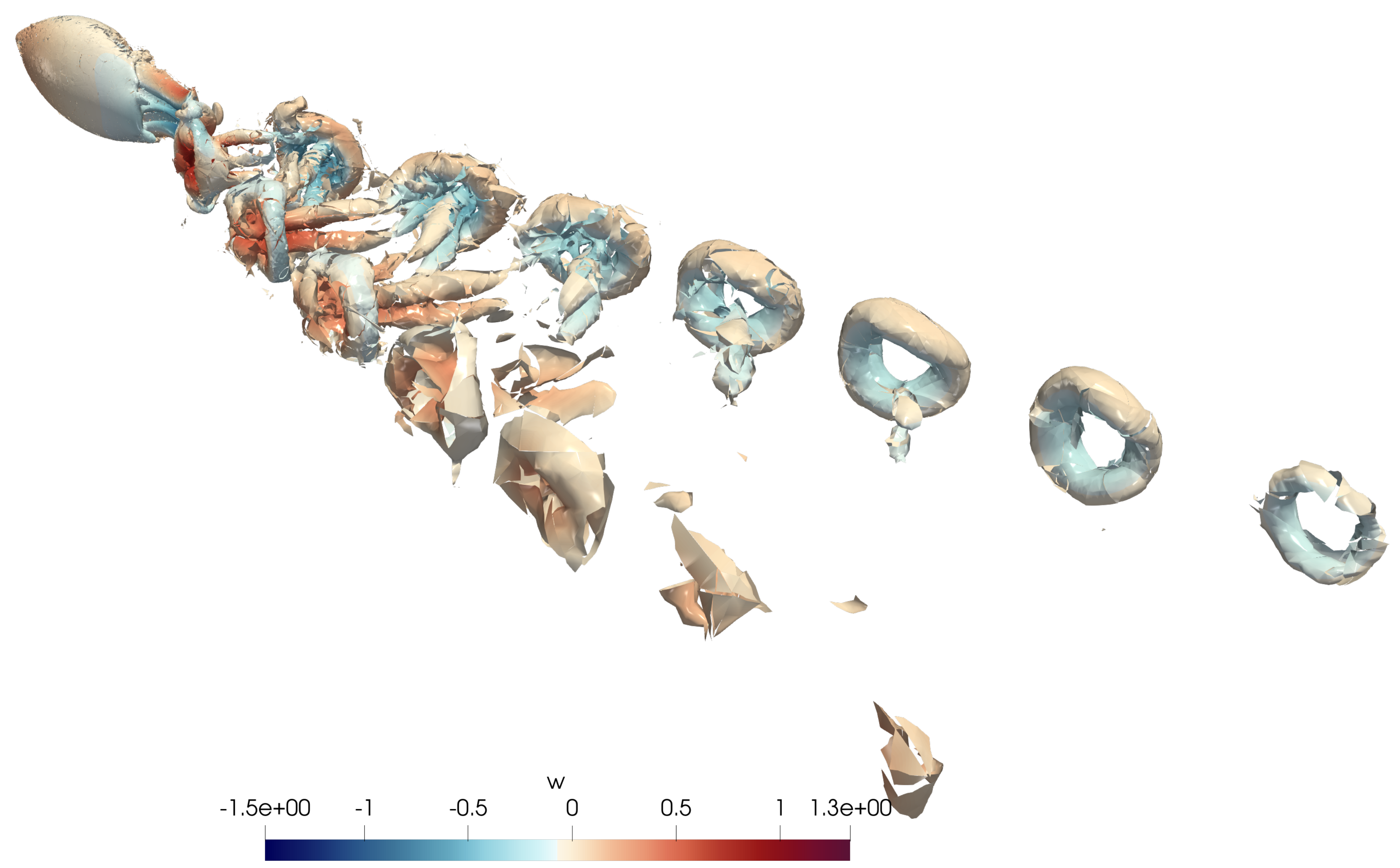}
        \caption{$Re=2400$, $t/T=0$}
    \label{fig:fish_2400_1}    
        \end{subfigure}
        \begin{subfigure}[b]{0.48\textwidth}
         \includegraphics[width=\textwidth]{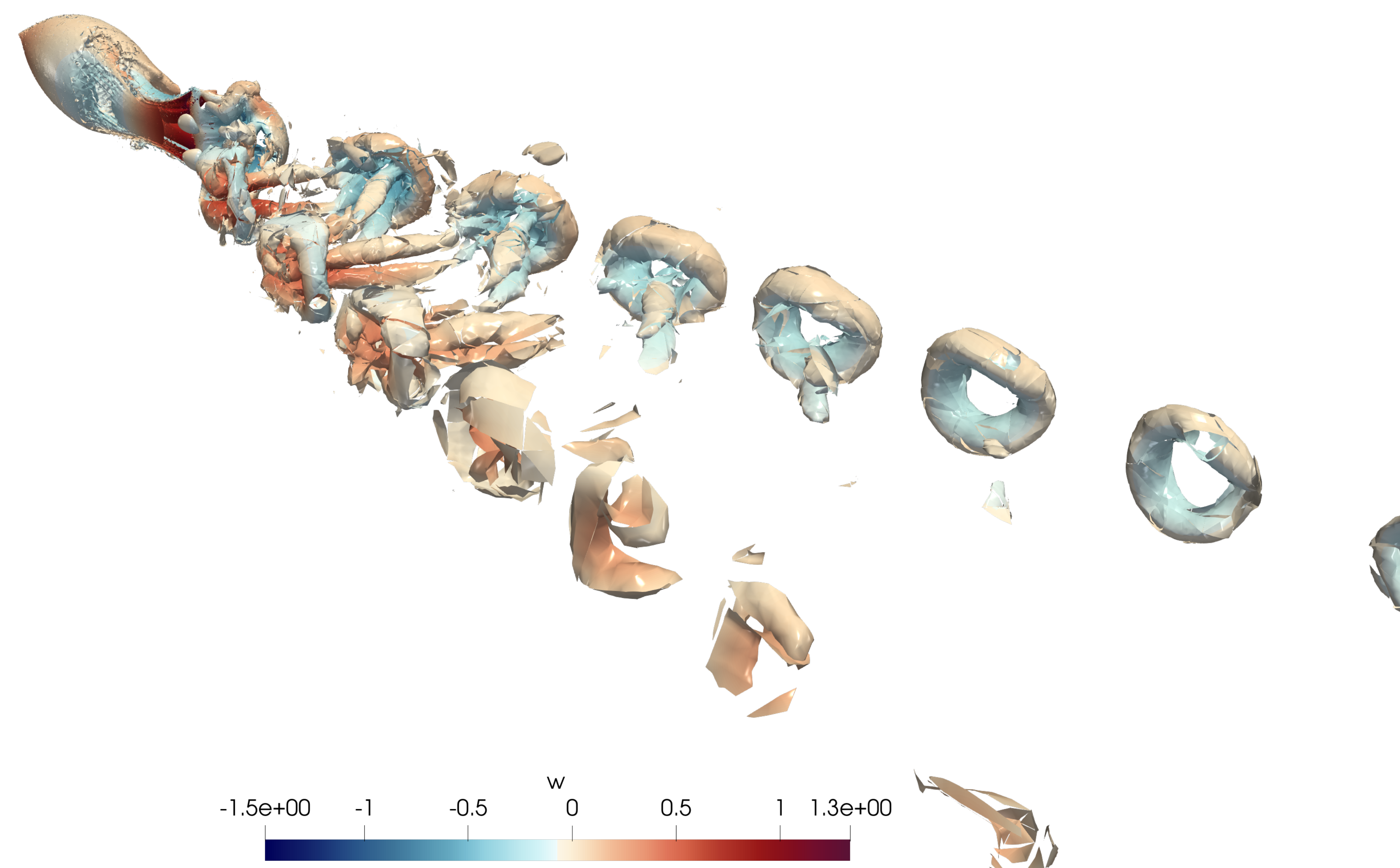}
        \caption{$Re=2400$, $t/T=1/2$}
    \label{fig:fish_2400_2}    
        \end{subfigure}
        \begin{subfigure}[b]{0.48\textwidth}
         \includegraphics[width=\textwidth]{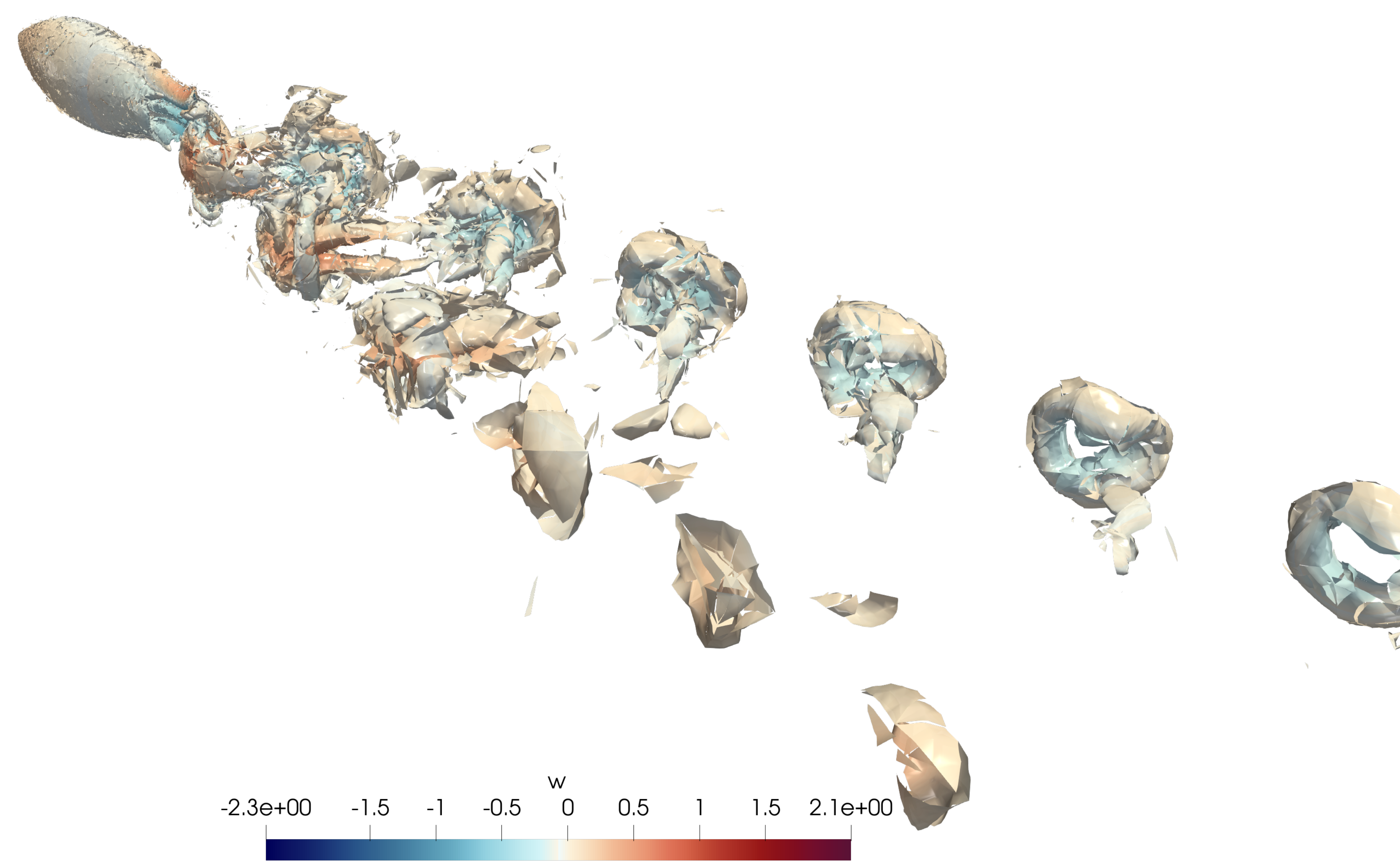}
        \caption{$Re=5800$, $t/T=0$}
    \label{fig:fish_5800_1}    
        \end{subfigure}
        \begin{subfigure}[b]{0.48\textwidth}
         \includegraphics[width=\textwidth]{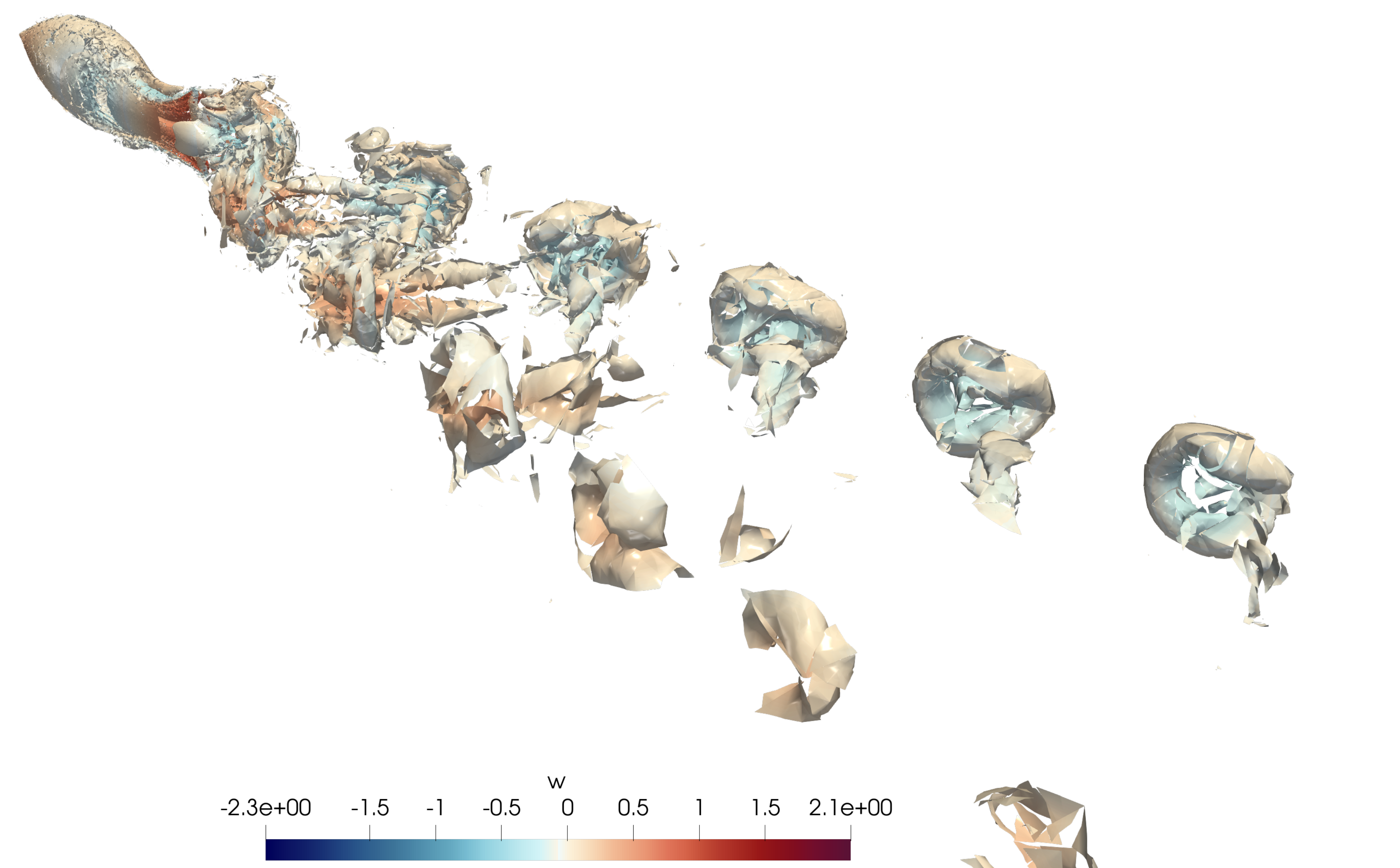}
        \caption{$Re=5800$, $t/T=1/2$}
    \label{fig:fish_5800_2}    
        \end{subfigure}
        \caption{Vortex structures behind the carangiform fish. The vortical structures are visualized with iso-surface of Q-criterion at $Q=0.02$.}
    \label{fig:fish}   
\end{figure}

The wake region of the carangiform fish for $Re=2400$ and $Re=5800$ is shown in Figure \ref{fig:fish}. The vortical structures are visualized using the second invariant of the velocity gradient defined as Q-criterion $Q = \frac12 \left(||\boldsymbol{\Omega}||^2 - ||\boldsymbol{S}||^2\right)$, where $\boldsymbol{S}$ and $\boldsymbol{\Omega}$ are symmetric and anti-symmetric components of the velocity gradient. The ring type vortical structures are observed in the wake region starting from the caudal fin. As these structures are convected in the flow direction, they organize into a staggered array of interconnected vortex rings. Notably, at the higher Reynolds number $Re=5800$, the wake exhibits a faster transition to finer, more chaotic scales compared to $Re=2400$. While the far-wake region is not fully resolved down to the smallest turbulent scales, the clean shedding and initial roll-up of the primary vortex rings demonstrate the robustness of the high-order ALE-DG method in handling complex, large-amplitude boundary deformations without suffering from grid-induced instabilities.

%% file: conclusion.tex
\label{sec:conclusion}
We have presented an ALE form of the Boltzmann equations, discretized with high-order discontinuous Galerkin method and semi-analytic Runge-Kutta methods with simplex elements. The resulting equations are used to simulate flows in nearly incompressible regime. The mesh movement is performed such that the elements remain straight sided and affine mapping is preserved. The geometric conservation law is satisfied by consistent updates of geometric factors with the mesh movement. The numerical results show that the geometric conservation law is satisfied in a free stream preservation test. The presented form is implemented for multi-GPU systems and tested with several numerical test cases. The results are satisfactory to model moving boundary problems in nearly incompressible regime with Galerkin-Boltzmann equations. The future work will include efficient mesh movement algorithms for complex boundary motions and fluid-structure interactions in nearly incompressible regime.